\documentclass[%
 aip,
amsmath,amssymb,
preprint,%
]{revtex4-1}

\usepackage{tabularx}
\usepackage[english]{babel}
\usepackage[utf8]{inputenc}
\usepackage[T1]{fontenc}
\usepackage{textcomp}
\usepackage{float}
\usepackage{amssymb,amsbsy} 
\usepackage{amsmath}
\usepackage{verbatim}
\usepackage{subfigure}
\usepackage{hyperref}
\usepackage{xcolor}
\usepackage{soul}
\usepackage{cleveref}
\usepackage{ragged2e}
\usepackage{fbox}
\usepackage{graphicx}
\usepackage{dcolumn}
\usepackage{bm}
\usepackage{dcolumn}
\usepackage{hyperref}
\usepackage[mathlines]{lineno}

\begin{document}



\title{On the Dual-Phase-Lag thermal response in the Pulsed Photoacoustic effect: a theoretical and experimental 1D-approach}

\author{L. F. Escamilla-Herrera}
\affiliation{Divisi\'on de Ciencias e Ingenier\'ias Campus Le\'on, Universidad de Guanajuato, A.P. E-143, C.P. 37150, Le\'on, Guanajuato, M\'exico.}
\author{J.M. Derramadero-Domínguez}
\affiliation{Departamento de Ingeniería Mecatrónica, Tecnológico Nacional de México / Instituto Tecnológico de Celaya, Av. Antonio García Cubas 600, Celaya, Gto., 38010, México.}
\author{O. M. Medina-Cázares} 
\affiliation{Divisi\'on de Ciencias e Ingenier\'ias Campus Le\'on, Universidad de Guanajuato, A.P. E-143, C.P. 37150, Le\'on, Guanajuato, M\'exico.}
\author{J. E. Alba-Rosales}
\affiliation{Centro de Investigaciones en Óptica AC, Loma del Bosque 115, CP 37150, León, GTO., México.}
\author{F.J. García-Rodríguez} 
\affiliation{Departamento de Ingeniería Mecatrónica, Tecnológico Nacional de México / Instituto Tecnológico de Celaya, Av. Antonio García Cubas 600, Celaya, Gto., 38010, México.}
\author{G. Gutiérrez-Juárez}\email{ggutj@fisica.ugto.mx}
\affiliation{Divisi\'on de Ciencias e Ingenier\'ias Campus Le\'on, Universidad de Guanajuato, A.P. E-143, C.P. 37150, Le\'on, Guanajuato, M\'exico.}

\date{\today}





\date{\today}
             
\begin{abstract}
In a recent work, assuming a Beer-Lambert optical absorption and a Gaussian laser time profile, the exact solutions for a 1D-photoacoustic(PA)-boundary value problem predict a null pressure for optically strong absorbent materials. To overcome this, a heuristic correction was introduced by assuming that heat flux travels a characteristic length during the duration of the laser pulse\cite{Ruiz-Veloz2021} $\tau_p$. In this work, we obtained exact solutions in the frequency domain for a 1D-boundary-value-problem for the Dual-Phase-Lag (DPL) heat equation coupled with a 1D PA-boundary-value-problem via the wave-equation. Temperature and pressure solutions were studied by assuming that the sample and its surroundings have a similar characteristic thermal lag response time $\tau_{_T}$, which was assumed to be a free parameter that can be adjusted to reproduce experimental results. Solutions for temperature and pressure were obtained for a three-layer 1D system. It was found that for  $\tau_{_T}< \tau_{p}$, the DPL temperature has a similar thermal profile of the Fourier heat equation, however, when $\tau_{_T}\ge \tau_{p}$ this profile is very different from the Fourier case. Additionally, via a numerical Fourier transform the wave-like behavior of DPL temperature is explored, and it was found that as $\tau_{_T}$ increases, thermal wave amplitude is increasingly attenuated.  Exact solutions for pressure were compared with experimental signals, showing a close resemblance between both data sets, particularly in the time domain, for an appropriated value of $\tau_{_T}$; the transference function was also calculated, which allowed us to find the maximum response in frequency for the considered experimental setup.

\end{abstract}

\maketitle




\section{\label{sec:I} Introduction}
Laser-induced ultrasound (LIU) longitudinal waves or pulsed photoacoustic (PPA) effect is the physical phenomenon in which a transference of fluence of a short laser pulse into condensed matter, generates a non-radiative transient producing stress waves usually in the ultrasound range, that propagates into the surrounding medium. Mechanical stress waves produced by LIU, are known to be strongly dependent on the features of the optical laser source, producing both longitudinal and transversal waves along with Rayleigh and Lamb waves  \cite{Minghua2006}.  Theoretical study of PPA effect and its generation techniques had a major resurgence in recent years, thanks to its wide range of applicability, for instance, the possible biomedical applications, such as PA imaging \cite{zhu2024,Jiang2024,Minghua2006,Chaigne2016}, or as a monitoring method in thermo-therapy for cancer treatment due how PPA allows to recognize optical absorbers in biological tissue \cite{Cox2009,Huang2013}.

In the thermoelastic regime, generation and propagation of PPA effect is described by the heat diffusion equation for temperature $T(\mathbf{r},t)$,  coupled with the wave equation for pressure, $P(\mathbf{r},t) $\cite{Cox2009,1968MorseIngard,1991Diebold}.
\begin{subequations}\label{eq1}
\begin{align}
&\left(\frac{\partial}{\partial t} -\chi \nabla^2\right) T(\mathbf{\mathbf{r}},t) = \frac{1}{\rho C_P}\left[ H(\mathbf{\mathbf{r}},t) + T_0\beta \frac{\partial}{\partial t} P(\mathbf{\mathbf{r}},t)\right]; \label{eq1a}\\ 
&\left(\rho_0\kappa_{_T}\frac{\partial^2}{\partial t^2}-\nabla^2\right) P(\mathbf{\mathbf{r}},t) = \rho_0\beta \frac{\partial^2}{\partial t^2} T(\mathbf{\mathbf{r}},t). \label{eq1b}
\end{align}
\end{subequations}

Where $\chi\equiv \kappa/\rho_0 C_P$ is the thermal diffusivity, $\kappa$ is thermal conductivity, $\rho_0$ and $T_0$ are the unaltered mass density and temperature, $C_P$ is the heat capacity at constant pressure, $\beta$ is the thermal coefficient of volume expansion, and, $\kappa_{_T}$ is the isothermal coefficient compressibility. $H(\mathbf{r},t)$ is the density of optical energy absorbed in the sample per unit-time also known as heating function. For material with low or moderate optical absorption, Eqs. \eqref{eq1}, can be uncoupled by invoking two approximations: the so-called stress (SC) and thermal confinements (TC), respectively, \cite{Cox2009, Ruiz-Veloz2021}.

The SC approximation implies that $dV(\mathbf{r},t)=[\beta T(\mathbf{r},t)-\kappa_{_T} P(\mathbf{r},t)]V_0\approx0$, \cite{Wang2007}, where $dV(\mathbf{r},t)$ is the fractional volume expansion and $V_0$ is the initial volume; i.e., volume remains constant during the heating process. On the other hand, the TC can be expressed as $-\nabla\cdot\nabla T(\mathbf{r},t)\approx0$, which means that the heat diffusion is negligible. If SC and TC are fulfilled, the temperature time derivative becomes proportional to $H(\mathbf{r},t)$ and, Eq. \eqref{eq1b} becomes the PA wave equation \cite{Ruiz-Veloz2021}, which is the cornerstone of PA imaging. However, both the SC and TC approximations predict at least two physical phenomena not considered in Eqs. \eqref{eq1}. First, with the SC approximation thermal waves or second sound phenomena are predicted via the relationship $T(\mathbf{r},t) \approx (\kappa_{_T}/\beta) P(\mathbf{r},t)$, which is no modeled by Eq. \eqref{eq1a}. Second, when TC approximation is considered the heat diffusion becomes a stationary phenomenon, valid just for low or moderate optical absorption materials \cite{Ruiz-Veloz2021}. Moreover, the PA wave equation, obtained when the SC and TC approximation are applied, is not well suited for any system, for instance, in optical strongly absorbent materials nor for systems in the mesoscopic scale\cite{Zhang2020}, although, they have been previously used for metals \cite{Sethuraman2007,Su2010}. Therefore, 
the PA model must be thoroughly analyzed and modified as consequence. 

Fourier heat conduction (FHC) law is the starting point in the PA model, which states that the rate of heat transfer passing through any material is proportional to the negative gradient of temperature; due to its simplicity and experimental accuracy, it is the most commonly used model to describe heat conduction \cite{Carslaw1959}. However, it also presents several physical inconsistencies; for instance, it is valid only under the assumption of local thermodynamic equilibrium, which is not fulfilled for short timescales and for very small systems \cite{Jou2010}. FHC law also predicts an infinite velocity for heat propagation, which poses a violation of causality, in such way that, when a change in temperature appears in any part of the system, an instantaneous perturbation also appears at each point of this system, i.e., a gradient in temperature at any given time $t_i$ in any point of the system, implies the appearing of an instantaneous heat flux everywhere in the system \cite{Joseph1989,OrdonezMiranda2009,Auriault2016}.

To overcome these inconsistencies, generalizations to Fourier heat conduction model have been proposed through the years, the simplest one is known as Maxwell-Cattaneo-Vernotte (MCV) heat conduction model \cite{Maxwell1866,Cattaneo1948,Vernotte1958} where a correction to the Fourier diffusion law is proposed by introducing a relaxation time for heat $\tau_q$ (also known as heat lag time) \cite{Auriault2016}, which is responsible for the so-called thermal inertial effect \cite{Christov2009, Straughan2011}, which is a heat wave phenomena. Similarly, the Dual-Phase-Lag (DPL) heat conduction model \cite{Tzou1997, Tzou2011}, proposes a similar constitutive relation between heat and temperature; however it is also modified by introducing yet another relaxation time $\tau_{_T}$, corresponding to a lagging in temperature (therefore, it is known as thermal lag time) \cite{Tzou1995}; this is why this model is known as Dual-Phase-Lag \cite{Rukolaine2014}. This second lag time is responsible for a non-stationary temperature propagation. This implies that thermal wave phenomenon and the non-stationary temperature propagation can be introduced in the PA model non-heuristically  via the DPL heat equation.

This work is organized as follows. In section \ref{sec:II}, we present a brief review on the different heat conduction models, and their constitutive relations, main features and some of the controversies regarding each one of them. With the aid of the thermal energy conservation equation, a modified PPA model based on the DPL heat conduction model will be considered to study. Section \ref{sec:III} presents a 1D three-layer photothermal boundary value problem with a PPA source produced by a pulsed laser with a Gaussian temporal width of $\sqrt{2}\tau_p$. Section \ref{sec:IV} presents analytical solutions for the DPL 1D heat equation in the frequency domain; via a numerical inverse Fourier transform, the corresponding thermal waves in time domain are also presented. These solutions are considered as the thermal photoacoustic source, and in section \ref{sec:V} the acoustic pressure solutions in 1D for this boundary problem are presented, considering thermal lag time $\tau_{_T}$ as a free parameter that can be set in order to better adjust theoretical results with experimental data. In section \ref{sec:VI} an analysis is presented, comparing the modified PPA 1D model with experimental data. Finally, conclusions and future perspectives of this research are given in section \ref{sec:VII}.


\section{\label{sec:II} Heat conduction models}


\subsection{Fourier heat conduction model}
 The corresponding constitutive relation between heat flux $q(\mathbf{r},t)$ and temperature $T(\mathbf{r},t)$ is given by \cite{Fourier1822}, $q(\mathbf{r},t) = -\kappa \nabla T(\mathbf{r},t)$.
If this relation is combined with the general equation for thermal energy conservation  in a certain domain $\Omega\in\mathbb{R}^n$ \cite{CalvoSchwarzwälder2015,Houssem2023},
\begin{equation}\label{eqconservation}
    C_P \rho_0\frac{\partial}{\partial t}  T(\mathbf{r},t)+ \kappa \nabla \cdot q(\mathbf{r},t) = H(\mathbf{r},t)\,;
\end{equation}
heat equation (\ref{eq1}a) is obtained, wich will be denoted in this work as parabolic heat equation (PHE). As mentioned in the introduction section, validity of Eq. \eqref{eqconservation} is limited to systems under local thermal equilibrium \cite{Rukolaine2014,OrdonezMiranda2009, Auriault2016}. 

\subsection{Maxwell-Cattaneo-Vernotte model}

The simplest generalization to FHC law is  introduced  by including a lag time $\tau_q$ on the heat flux in order to overcome the infinity heat propagation speed. The corresponding constitutive relation for heat and temperature of the MCV heat conduction model is then given by, $q(\mathbf{r},t+\tau_q) = -\kappa \nabla T(\mathbf{r},t)$; if a Taylor series expansion up to first-order expansion in $\tau_q$ is performed, the MCV approximated constitutive relation is then given by,
\begin{equation}\label{thermalinertia}
    \left(1 + \tau_q \frac{\partial}{\partial t} \right) q(\mathbf{r},t) = -\kappa \nabla T(\mathbf{r},t)\,.
\end{equation}

The second term left-hand-side is referred in the literature as \textit{thermal inertia}, since it is the intensive property of materials or substances to store heat, retain it and then release over a span of time as an analogy of its mechanical counterpart; this property is also related with heat capacity and thermal conductivity of materials \cite{Joseph1989,Holba2015}. When Eq. \eqref{thermalinertia} is combined with thermal energy conservation given by Eq. \eqref{eqconservation}, leads to the following hyperbolic heat equation (HHE), 
\begin{equation}\label{Heat_transport_MCV}
\left[ \frac{1}{\Tilde{\chi}}\left( \frac{\partial}{ \partial t} +\tau_{q} \frac{\partial^2}{ \partial t^2}\right)-\nabla^2 \right]T(\mathbf{r},t) = \frac{1}{ \kappa} \left(1+\tau_{q}\frac{\partial}{\partial t} \right)H(\mathbf{r},t)\,.
\end{equation}

The main feature of Eq. \eqref{Heat_transport_MCV} is the fact that it indeed avoids the paradox of infinite heat speed, related to the FHC model \cite{Ozisik1994,OrdonezMiranda2009,Auriault2016}. This is the reason why several authors have proposed that MCV heat conduction model is able to extend the FHC regime to timescales shorter than the heat relaxation time of the system under study. Another important feature of this equation is the appearance of the quantity $\tau_q/\Tilde{\chi}$ which has speed dimensions; this term is known as \textit{second sound}, which refers to the propagation speed of the temperature field as thermal waves, in analogy to the pressure field (or mechanical lattice vibrations) propagation as acoustic waves \cite{Beardo2021}. The thermal relaxation time $\tau_q$ could be then understood as a phase lag of heat, needed to reach the steady heat state in the system when the temperature suddenly changes.

Theoretical foundations of second sound for fluids were first given in the 1950's \cite{Ward1952,London1954}; and in the context of solids during the 1960's \cite{Chester1963,Enz1968,Hardy1970}. Without giving much detail, in these works, second sound, or thermal waves are microscopically modeled as a coherent propagation of density perturbations in the phonon gas (therefore, thermal waves are a non-equilibrium  phenomena), and heat lag time can be calculated for solids in terms of bulk parameters as $\tau_q = 3 \kappa_{ij}/ a C_P$, where $\kappa_{ij}$ is the thermal conductivity tensor and $a$ is the Young's modulus \cite{Auriault2016}. For pure, homogeneous and thermal conductor materials $\tau_q$ has been calculated to be in the range of pico-seconds. Recently, in Ref. \cite{Beardo2021}, heat lag time is reported to be measured for Ge at room temperature, up to our knowledge this results have not been reproduced yet by others.    

However, the MCV heat conduction model is not free of inconsistencies, for instance, in Ref. \cite{Bai1995}, HHE is reported to give results which are not consistent with the second law of thermodynamics, in the context of non-equilibrium rational thermodynamics due to the appearance of negative solutions for temperature in absolute scales; additionally, still there are other theoretical concerns about the use of HHE as a generalization of PHE, in particular, regarding thermodynamics and statistical mechanics perspective \cite{Bright2009}.

\subsection{Dual-Phase-Lag model}

The second generalization to FHC model considered in this work, is known in the literature as Dual-Phase-Lag model \cite{Tzou1997, Tzou2011} which considers not only the heat lag time $\tau_q$ as in the MCV case, but also a thermal relaxation time (or thermalization time) $\tau_{_T}$, known as thermal lag time; therefore, this model introduces two different lagging phenomena for heat conduction. Heat lag time $\tau_{_q}$, has a response in the short-times scale, while $\tau_{_T}$ considers the small spatial scale. Due to this last feature of the DPL model and in particular for strongly optical absorbent materials, the thermal source occurs around of 1 optical path length (of the order of nm).

Constitutive relation corrected with a dual-phase lag (DPL) in heat flux and thermal response is then given by $ q(\mathbf{r},t+\tau_q) = -\kappa \nabla T(\mathbf{r},t+\tau_{_T})$.
A double Taylor series expansion up to first order in time is performed both in $\tau_q$ and in $\tau_{_T}$, obtaining the DPL constitutive relation as a first-order perturbation expansion for both lag times,
\begin{equation}\label{DPLexpansion}
    \left(1 + \tau_q \frac{\partial}{\partial 
     t}\right)q(\mathbf{r},t) = -\kappa \left(1 + \tau_{_T} \frac{\partial}{\partial t}\right) \nabla T(\mathbf{r},t)\,.
\end{equation}
This equation reduces to the MCV case presented in Eq. \eqref{thermalinertia} as  $\tau_{_T}\to0$ and then to the well known Fourier's law of heat conduction given if $\tau_q\to0$.

Combining Eq. \eqref{DPLexpansion}  with Eq. \eqref{eqconservation}, leads to the following differential equation for the Dual-Phase-Lag heat conduction with a thermal source term $H(\mathbf{r},t)$,
\begin{equation}\label{Heat_transport_DPL}
\left[ \frac{1}{ \Tilde{\chi}}\left( \frac{\partial}{\partial t} +\tau_{q} \frac{\partial^2}{\partial t^2}\right)-\left(1+\tau_{_T} \frac{\partial}{\partial t}\right)\nabla^2 \right]T(\mathbf{r},t) =\frac{1}{\kappa} \left(1+\tau_{q} \frac{\partial}{ \partial t} \right)H(\mathbf{r},t)\,.
\end{equation} 

In this work, Eq. \eqref{Heat_transport_DPL} will be considered to model heat propagation in the PPA effect instead of the usual FHC model given by Eq. (\ref{eq1}a), for a 1D boundary three-layer problem; additionally, we also consider that: (a) the SC is fulfilled, which is implemented in order to allow us to mathematically uncouple both DPL and acoustic differential equations, where $\Tilde{\chi} \equiv [1 - (T_0 \beta^2/C_p \rho_0 \kappa_{_T})]^{-1} \chi $ which is the modified thermal diffusivity corrected by stress confinement; however, for most materials it can be noticed that $\Tilde{\chi}\approx\chi$; (b) the generation and propagation of the temperature, is therefore governed by the DPL equation \eqref{eqconservation} and; (c) their solutions are the source of the PA wave equation Eq. (\ref{eq1}b).




\section{\label{sec:III} 1D photothermal boundary value problem in the time domain}

\begin{figure}[H]
\centering
\includegraphics[width=0.4\textwidth]{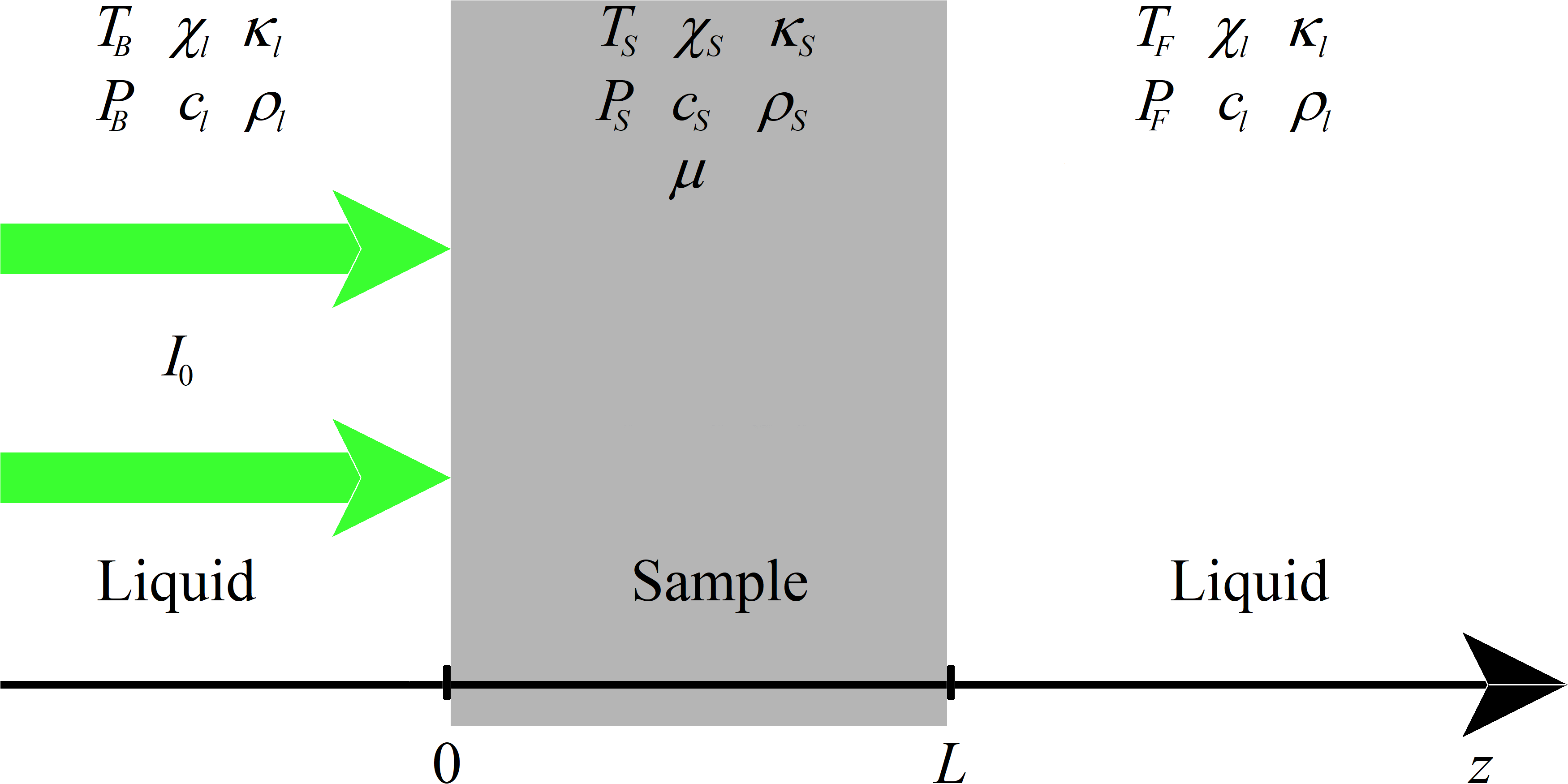}\\
\hrulefill \\
\caption{Geometry of the 1D three-layer system. A semi-infinite slab of width $L$ made of a strongly optical absorbent material sample is surrounded by non-absorbent fluid in the B and F regions. A pulsed laser then illuminates the backward face on the slab, producing a heating on the sample modeled by Eq. \eqref{Heat_transport_DPL} ; as a response, the sample produces longitudinal acoustic waves propagating to both backward and forward regions.}
\label{3layer}
\end{figure}

In order to explore how the PPA model modified by substituting the usual heat diffusion equation with the DPL heat equation given by Eq. \eqref{Heat_transport_DPL} alters the acoustic waves induced, in this section, a 1D three-layer system is under study. The geometry of this problem is presented in Fig. \ref{3layer}, a semi-infinite slab (S) of a strongly optical absorbent material sample of width $L$ is located at $z = 0$; it is surrounded  by a non-absorbent fluid; the interval $(-\infty,0)$ is called backward (B) region and the interval $(L,\infty)$ is called forward region (F).

 As a first approach, in this work we have considered a system for which thermal lag time $\tau_{_T}$ is considered to be the same regardless of the region or material. Under this assumption, heat conduction is then given by,
\begin{subequations}\label{1DTemperature}
\begin{align}
&\left[ \frac{1}{ \tilde{\chi}}_l\left( \frac{\partial}{ \partial t} + \tau_{q_l}\frac{\partial ^2}{\partial t^2}\right)-\left(1+\tau_{_T} \frac{\partial}{\partial t} \right) \frac{ \partial^2}{\partial z^2}\right]T_B(z,t) = 0;  \hspace{0.2\linewidth}z<0  \\ 
&\left[ \frac{1}{ \tilde{\chi}}_s\left( \frac{\partial}{ \partial t} + \tau_{q_s}\frac{\partial ^2}{\partial t^2}\right)-\left(1+\tau_{_T} \frac{\partial}{\partial t} \right) \frac{ \partial^2}{\partial z^2}\right]T_S(z,t) = \frac{1}{\kappa_s} \left(1+\tau_{q_s} \frac{\partial}{\partial t}\right)H(z,t); \ 0\le z \le L\\ 
&\left[ \frac{1}{ \tilde{\chi}}_l\left( \frac{\partial}{ \partial t} + \tau_{q_l}\frac{\partial ^2}{\partial t^2}\right)-\left(1+\tau_{_T} \frac{\partial}{\partial t} \right) \frac{ \partial^2}{\partial z^2}\right]T_F(z,t) = 0 \hspace{0.2\linewidth} L<z  \,.
\end{align}
\end{subequations}
Where indexes $l$ and $s$ stands for the corresponding values in the fluid and the slab, respectively. These equations are coupled with the pressure equations for each of the three layers,
\begin{subequations}\label{1DPressure}
\begin{align}
\left[\frac{1}{c_l^2} \frac{\partial^2}{ \partial t^2} - \frac{\partial^2}{\partial z^2}\right]P_B(z,t) &= \beta_l\rho_l \ \frac{\partial^2}{\partial t^2} T_B(z,t);  \hspace{0.01\linewidth} z<0  \\ 
\left[\frac{1}{c_s^2} \frac{\partial^2}{ \partial t^2} - \frac{\partial^2}{\partial z^2}\right]P_S(z,t) &= \beta_s\rho_s \ \frac{\partial^2}{\partial t^2} T_S(z,t); \hspace{0.01\linewidth} 0\le z \le L\\ 
\left[\frac{1}{c_l^2} \frac{\partial^2}{\partial t^2} - \frac{\partial^2}{\partial z^2}\right]P_F(z,t) &= \beta_l\rho_l \ \frac{\partial^2}{ \partial t^2} T_F(z,t) \hspace{0.01\linewidth}; L<z  \,.
 \end{align}
\end{subequations}


In order to solve Eqs. \eqref{1DTemperature} and \eqref{1DPressure}, the photo-thermal Cauchy boundary conditions (BC) for the temperature and pressure are considered \cite{Ruiz-Veloz2021}. These conditions are explicitly presented in Table \ref{boundary} of the supplemental material. In addition to the boundary conditions, the following restrictions must be imposed: (a) $\lim_{z\to-\infty} T_{B}(z,t) = \lim_{z\to\infty} T_{F}(z,t) =0\,,$ to take into account the fact that temperature is strongly attenuated far away from the thermal source (slab) in any direction. Additionally, regarding acoustic pressure, we will assume that in the B region, there will be only acoustic waves moving towards $z\to-\infty$, and conversely, in the F region, there will be only acoustic waves moving towards $z\to\infty$.


$H(z,t)$ is defined to be the density of electromagnetic energy absorbed by the sample, per unit of time \cite{Ruiz-Veloz2021}. We are interested in the case of an uniformly illuminated flat-slab sample, for which the spatial electromagnetic absorption rate $h(z)$ is modeled by the Beer-Lambert law \cite{Attard1998,fox2010}, $h(z) = I_0 \mu e^{-\mu|z|}$ where, $I_0$ is the laser fluence $(J/m^2)$ and $\mu$ is the optical absorption coefficient of the sample. With  a Gaussian pulse with a 1/e width (duration) of $\sqrt{2}\tau_p$  the heating function in time domain is then written as,
\begin{equation}\label{Heating}
    H(z,t) = h(z) \cdot \Theta(t) =  I_0 \mu e^{-\mu|z|} \frac{2}{\tau_p \sqrt{\pi}}\exp\left[-2\left(\frac{t}{\tau_p}\right)^2\right]\,.
\end{equation}

Temperature and pressure partial differential equations for the 1D photothermoacoustic boundary value problem given by Eqs. \eqref{1DTemperature} and \eqref{1DPressure} are not straightforward to solve in time domain \cite{Lam2013}; in order to have analytical solutions for the aforementioned system of equations, we have considered to work in the frequency domain instead, given how these partial differential equations are simplified in this space.



\section{\label{sec:IV} 1D temperature solutions in the frequency domain}

Before proceeding with the solution of the 1D photothermoacoustic problem with boundary conditions, it is necessary to remark the main assumption of this work, throughout this research the thermal lag response time, $\tau_{_T}$ in Eq. \eqref{1DTemperature}, will be consider as an independent physical variable, since to our knowledge there are not yet available first principle proposals on how calculate specific values for $\tau_{_T}$.
Therefore, in the following we will denote temperature as, $T_i = T_i(z,t;\tau_{_T})$, with $i=\{F,S,B\}$; similarly, the acoustic pressure will be consider to have the same functional dependency, $P_i = P_i(z,t;\tau_{_T})$.

Inverse Fourier transform is then applied to Eqs. \eqref{1DTemperature}, reducing the DPL differential equations for the 1D three-layer system in the frequency domain to,
\begin{subequations}\label{Temperature_frequency}
\begin{align}
&\left[\frac{\partial^2}{\partial z^2}-\frac{1}{\tilde{\chi_l}} \left(\frac{i\omega - \tau_{ql}\omega^2}{1+ i\tau_{_T}\omega}\right) \right]\hat{T}_B(z,\omega;\tau_{_T}) = 0,  \hspace{0.31\linewidth}z<0;  \\ 
&\left[\frac{\partial^2}{\partial z^2}-\frac{1}{\tilde{\chi_s}}\left(\frac{i\omega - \tau_{qs}\omega^2}{1+ i\tau_{_T}\omega}\right) \right]\hat{T}_S(z,\omega;\tau_{_T}) = -\frac{1}{\kappa_s}\left(\frac{1 + i\tau_{qs}\omega}{1+ i\tau_{_T}\omega} \right) h(z)\hat{\Theta}(\omega),\hspace{0\linewidth}0\le z \le L;\\ 
&\left[\frac{\partial^2}{\partial z^2}-\frac{1}{\tilde{\chi_l}}\left(\frac{i\omega - \tau_{ql}\omega^2}{1+ i\tau_{_T}\omega}\right) \right]\hat{T}_F(z,\omega;\tau_{_T}) = 0; \hspace{0.3\linewidth} L<z.  \,
\end{align}
\end{subequations}
Where $\omega$ is the angular frequency, defined in terms of the frequency as $\omega = 2\pi f$.

Also, in Eqs. \eqref{Temperature_frequency} defined the DPL thermal wave number $\sigma^2_j (\omega;\tau_{_T})$,  for the $j$-th (with $j = l,s$) material as,
\begin{equation}\label{sigmaDPL}
    \sigma^2_j (\omega;\tau_{_T}) = \frac{1}{\Tilde{\chi}_j} \frac{i\omega - \tau_{qj} \omega^2}{1 + i\tau_{_T} \omega}\,.
\end{equation}

An analogous definition can be given for the PHE and HHE thermal wave numbers, as follow,
\begin{subequations}\label{sigmas}
\begin{align}
     \sigma^2_{j_\text{HHE}} (\omega;\tau_{_T}) &=  \frac{1}{\Tilde{\chi}_j}\left(i\omega - \tau_{qj} \omega^2 \right) ; \quad (\tau_{_T}\to0); \\
     \sigma^2_{j_\text{PHE}} (\omega;\tau_{_T}) &= i\frac{\omega}{\Tilde{\chi}_j}; \quad   (\tau_{_T}\to0 \ \text{and} \ \tau_{qj}\to0) \,.
\end{align}
\end{subequations}
Therefore, DPL thermal wave number given by Eq. \eqref{sigmaDPL}, is the most general expression for this quantity. 

For the sake of simplicity, in the following, the functional dependence of $(\omega;\tau_T))$ for the most functions will be omitted. Regarding the considered boundary conditions in frequency domain are not significantly modified, replacing $t$ for $\omega$ in the functional dependence of temperature and acoustic pressure.

Heating function in the frequency domain $H(z,\omega)$ is,
\begin{displaymath}
    H(z,\omega) = h(z)\cdot\hat{\Theta}(\omega) = I_0 \mu e^{-\mu|z|} \frac{1}{\sqrt{\pi}}\exp\left[-\frac{1}{2}\left(\frac{\tau_p \omega}{2}\right)^2\right]\,.
\end{displaymath}

The family of solutions of Eqs. \eqref{Temperature_frequency} for this problem are 
\begin{subequations}\label{Temperature_solutions}
\begin{align}
\hat{T}_B(z,\omega;\tau_{_T}) &= a_{1B} e^{\sigma_l z} + a_{2B} e^{-\sigma_l z};\hspace{0.34\linewidth} z < 0 \\
\hat{T}_S(z,\omega;\tau_{_T}) &= a_{1S} e^{\sigma_s z} + a_{2S} e^{-\sigma_s z} + \hat{T}_{0}(\omega;\tau_{_T})e^{-\mu z}; \hspace{0.15\linewidth} 0 < z < L \\
\hat{T}_F(z,\omega;\tau_{_T}) &= a_{1F} e^{\sigma_l z} + a_{2F} e^{-\sigma_l z}; \hspace{0.34\linewidth} L < z
\end{align}
\end{subequations}

where $a_{nm}(\omega;\tau_{_T})$ (with $n = \{1,2\}$ and $m = \{F,S,B\}$) are the undetermined coefficients on each layer and $\hat{T}_0(\omega;\tau_{_T})e^{-\mu z}$ is the particular solution of the non-homogeneous differential equation for the slab S,
\begin{equation}\label{ParticularDPL}
    \hat{T}_{0}(\omega;\tau_{_T}) = -\frac{I_{0 }\mu}{\kappa_{s} }\frac{  (1 + i \omega \tau_{_T})(1 + i \tau_{qs} \omega)}{ (\mu^2 - \sigma^2_{s})(1 + i \tau_{_T} \omega)} \ \hat{\Theta}(\omega)\,.
\end{equation}

As in the case of $\sigma_j$, the particular solutions for each one of the thermal conduction equations (i.e., HHE and PHE),are particular cases of Eq. \eqref{ParticularDPL}. Specifically, 
\begin{align*}
    \hat{T}_{0_\text{HHE}}(\omega) &= -\frac{I_{0 }\mu}{\kappa_{s} }  \left ( \frac{  1 + i \tau_{qs} \omega}{\mu^2 - \sigma^2_{s_\text{HHE}}}\right)\ \hat{\Theta}(\omega)\,; \quad \quad \quad\quad\quad(\tau_{_T}\to0); \\
    \hat{T}_{0_\text{PHE}}(\omega) &= -\frac{I_{0 }\mu}{\kappa_{s} } \left (\frac{1}{ \mu^2 - \sigma^2_{s_\text{PHE}}}\right) \ \hat{\Theta}(\omega)\,; \qquad \quad (\tau_{_T}\to0 \ \text{and} \ \tau_{qj}\to0) \,
\end{align*}

Applying boundary conditions for $T$ presented in supplemental material Table I in frequency domain and physical restrictions to the solutions for temperature, undetermined coefficients $a_{nm}(\omega;\tau_{_T})$ are found. These coefficients are presented in Eqs. (S.1.) of the supplemental material. The functional form of each of these coefficients is the same regardless the heat model (FHE, MCV or DPL) selected. 
Additionally, regarding these coefficients, it must be mentioned that in the regime of high frequencies (>1 MHz), it is observed that $a_{1S}\ll a_{2S}$; therefore, coefficient $a_{1S}$ can be disregarded without any loss of generality in temperature solutions given by Eqs. \eqref{Temperature_solutions}.



\section{\label{sec:V} 1D pressure solutions in the frequency domain}

Applying the inverse Fourier transform to Eqs. \eqref{1DPressure} we obtained the corresponding equations in frequency domain,
\begin{align*}
\left(\frac{\partial^2}{\partial z^2}-k^2_l\right)\hat{P}_B(z,\omega;\tau_{_T}) &= -\omega^2 \beta_l\rho_l \hat{T}_B(z,\omega;\tau_{_T}); \hspace{0.01\linewidth}  z<0  \\ 
\left(\frac{\partial^2}{\partial z^2}-k^2_s\right)\hat{P}_S(z,\omega;\tau_{_T}) &= -\omega^2 \beta_s\rho_s \hat{T}_S(z,\omega;\tau_{_T}); \hspace{0.01\linewidth} 0 \le z \le L \\
\left(\frac{\partial^2}{\partial z^2}-k^2_l\right)\hat{P}_F(z,\omega;\tau_{_T}) &= -\omega^2 \beta_l\rho_l \hat{T}_F(z,\omega;\tau_{_T});  \hspace{0.01\linewidth} L<z \,.
\end{align*}

Where we have defined the acoustic wave number $k_j=\omega/c_j$. Solutions of the above set of non-homogeneous differential equations are,
\begin{subequations}\label{Pressure_solutions}
\begin{align}
\hat{P}_B(z,\omega;\tau_{_T}) &= b_{1B} e^{+ik_l z} + b_{2B} e^{-ik_l z} + \hat{P}_{T_B}(z,\omega;\tau_{_T}); \hspace{0.16\linewidth}z < 0 \\
\hat{P}_S(z,\omega;\tau_{_T}) &= b_{1S} e^{+ik_s z} + b_{2S} e^{-ik_s z} + \hat{P}_{T_S}(z,\omega;\tau_{_T});\hspace{0.15\linewidth} 0 \le z \le L \\
\hat{P}_F(z,\omega;\tau_{_T}) &= b_{1F} e^{+ik_l z} + b_{2F} e^{-ik_l z}+ \hat{P}_{T_F}(z,\omega;\tau_{_T}); \hspace{0.15\linewidth} L < z
\,.
\end{align}
\end{subequations}
Where $b_{nm}(\omega;\tau_T)$ are the undetermined coefficients for the solutions of pressure equations; the set of particular solutions for pressure, $\hat{P}_{T_i}(z,\omega;\tau_{_T})$ is then given by,
\begin{subequations}\label{Pressure_particular}
\begin{align}
\hat{P}_{B0}(z,\omega;\tau_{_T}) &= \beta_{l} \rho_{l}\frac{\omega^2 }{k_{l}^2 + \sigma_{l}^2} a_{1B}(\omega;\tau_{_T})e^{+\sigma_{l} z} \,; \\
\hat{P}_{S0}(z,\omega;\tau_{_T}) &=   \beta_{s} \rho_{s} \frac{\omega^2 }{k_{s}^2 + \sigma_{s}^2}\left[e^{\mu z}a_{2S}(\omega;\tau_{_T})+\frac{\sigma_s^2 +\kappa_s^2}{\mu^2 + k_{s}^2} e^{\sigma_{s} z}  \hat{T}_{0}(\omega;\tau_{_T}) + \right] e^{- (\sigma_{s} + \mu)z}\,; \\
\hat{P}_{F0}(z,\omega;\tau_{_T}) &= \beta_{l} \rho_{l}\frac{\omega^2 }{k_{l}^2 + \sigma_{l}^2} a_{2F}(\omega;\tau_{_T}) e^{-\sigma_{l} z}\,.
\end{align}
\end{subequations}

Applying the boundary conditions for pressure given in Table I and the corresponding physical constrictions, the undetermined coefficients can be calculated; these coefficients can be explicitly found in Eqs. S.2 of the supplemental material.

%


\section{\label{sec:VI} Numerical calculations}

\subsection{DPL temperature}

The 1D three-layer system is modeled as follows: B and F regions are considered to be water and the slab sample S to be made of aluminum of width  $L = 3.4$ mm, the whole system is held at thermal equilibrium at room temperature (300 K) and at a pressure of 1 atm. A pulsed laser illuminates the face of the slab located at $z = 0$, producing a heating modeled by the Beer-Lambert law with a fluence  of $I_0 = 10$ mJ and a Gaussian temporal profile with $\tau_p = 10$ ns.  

\begin{figure}
  \centering
  \subfigure[]{
    \includegraphics[width=0.4\textwidth]{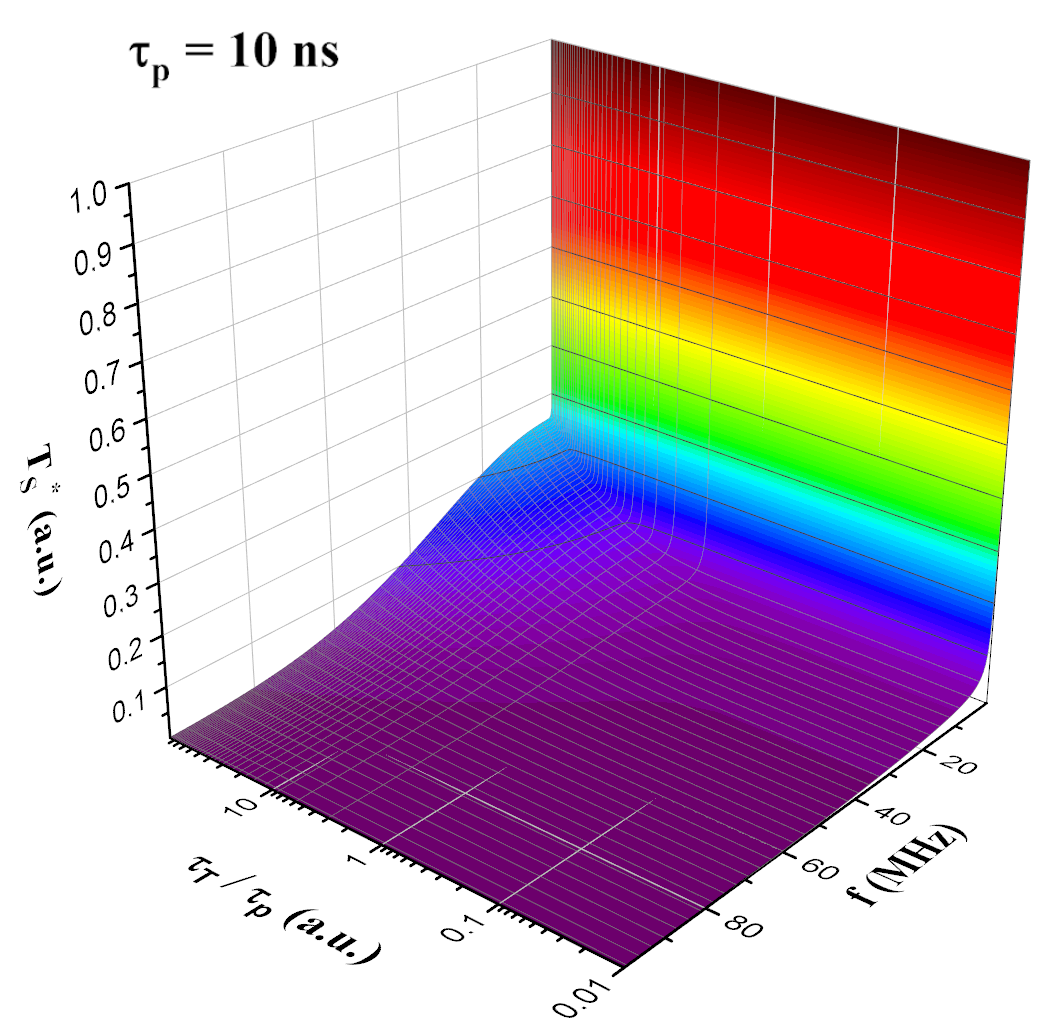}
    \label{fig:3DTemp}
  }
  \subfigure[]{
    \includegraphics[width=0.4\textwidth]{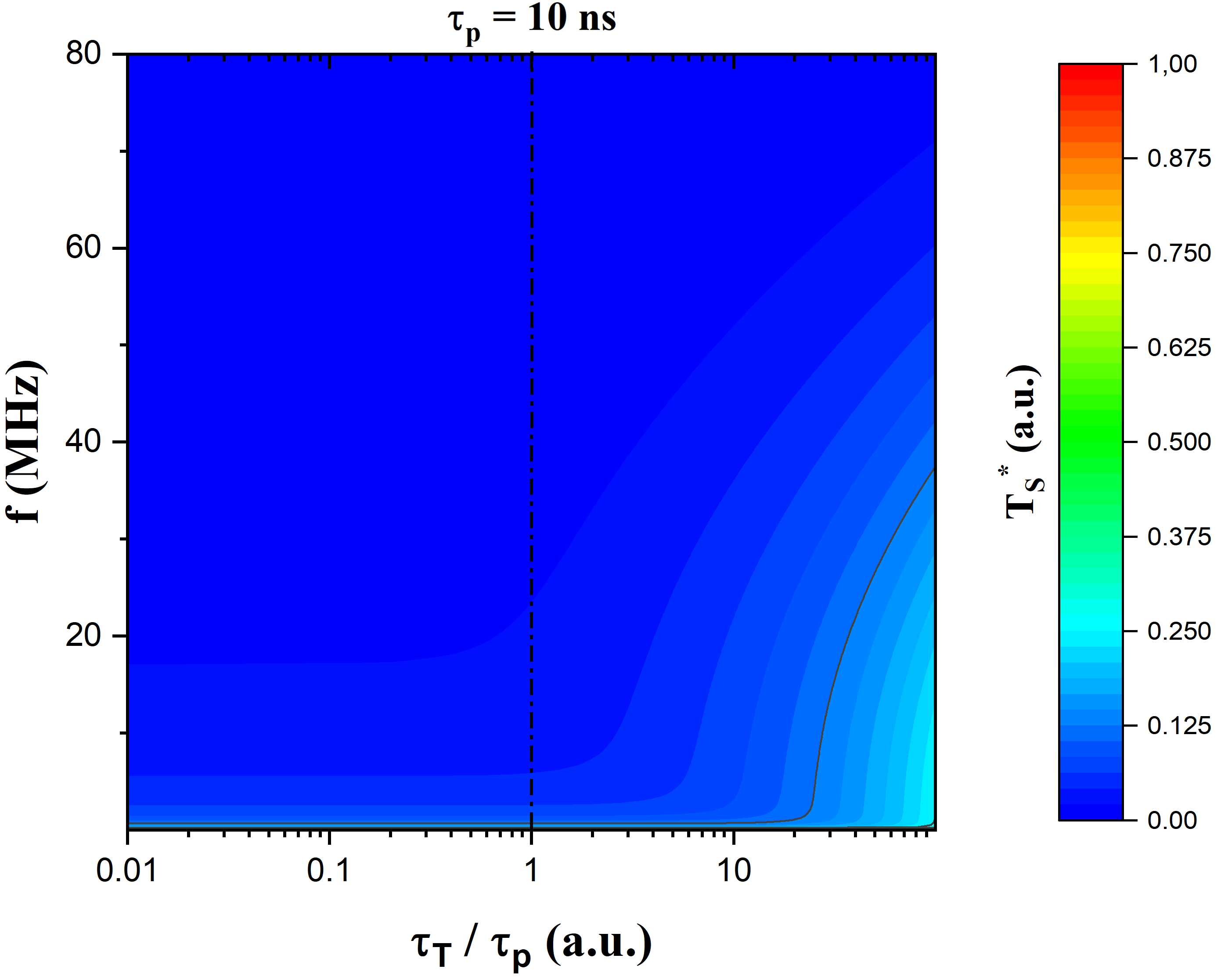}
    \label{fig:TempCon}
  }
  \\
  \hrulefill \\
  \caption{Normalized plots for the real part of the DPL heat transport equation solution presented in Eq. (\ref{Temperature_solutions}b) of the material slab $\hat{T}_S$, considering a sample composed by aluminum calculated at $z=0$ and a laser pulse time of $\tau_p = 10$ ns. In \subref{fig:3DTemp} 3D temperature plot is presented as a function of thermal lag time $\tau_{_T}$ and frequency $f$. In \subref{fig:TempCon} the corresponding contour plot is presented. The dash-dot line in the contour plot separates the regions for which $\tau_{_T}\approx\tau_p.$}
  \label{fig:fig2}
\end{figure}

Plots for temperature $\hat{T}_S (0,f;\tau_{_T})$ are presented in Fig. \ref{fig:fig2}, as a function of $\tau_{_T}$ and $f$. For such plots, temperature is normalized with respect to its maximum value and $\tau_{_T}$ is normalized with respect to $\tau_p$. In Fig. \ref{fig:3DTemp}, 3D plot for temperature $\hat{T}_S$ is presented, as noticed, for $\tau_{_T} < \tau_p$ temperature exhibits $f^{-1/2}$ decay. This behavior was previously observed for the PHE in frequency domain \cite{Ruiz-Veloz2021}. However, DPL heat equation has a different behavior in the region where $\tau_{_T} \approx \tau_p$, showing a less pronounced decreasing with frequency, being more significant as thermal lag time is greater compared with laser pulse time. In Fig. \ref{fig:TempCon} this behavior is more evident in the corresponding contour plot, where a line is set in order to separate the aforementioned regions. 

Therefore, it is evident that for a variable thermal lag time $\tau_{_T}$ in the 1D DPL heat equation, laser pulse time $\tau_p$ corresponds to a threshold in time, for which, smaller values of $\tau_{_T}$ (with respect to $\tau_p$) there are not significant changes when compared to the standard PHE model, however, once it is once surpassed its behavior is strongly modified. 

\begin{figure}[h]
\centering
\includegraphics[width=0.4\textwidth]{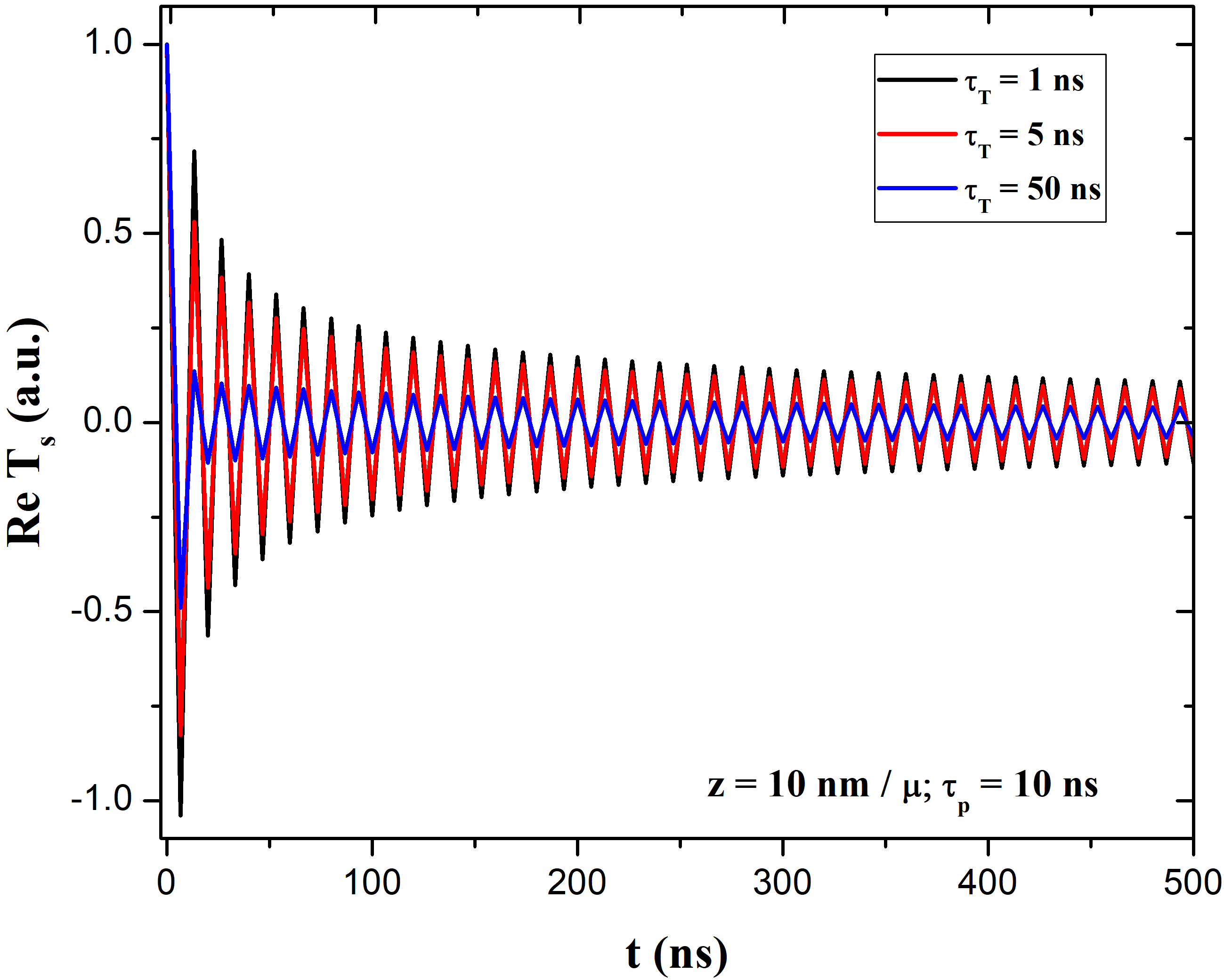}\\
\hrulefill \\
\caption{Numerical Fourier transform for normalized real part of the temperature in the aluminum slab as a function of time, calculated at a sample depth of $z = 10 nm/\mu$, for different values of $\tau_{_T}$.  Black curve corresponds to $\tau_{_T} = 1$ ns, the red one to $\tau_{_T} = 5$ ns and the blue curve to $\tau_{_T} = 50$ ns.}
\label{IFFT_Tslab}
\end{figure}

In order to explore further the effect of $\tau_{_T}$ on the temperature, a numerical inverse Fourier transformation is then applied to the DPL 1D temperature solution in the sample (\ref{Temperature_solutions}b), for three different values of $\tau_{_T}$, which are presented in Fig. \ref{IFFT_Tslab}, namely, $\tau_{_{T1}} = 1$ ns (black line), $\tau_{_{T2}} = 5$ ns (red line) and $\tau_{_{T3}} = 50$ ns (blue line). 

Temperature is calculated at a sample depth of 10 nm normalized with the optical penetration depth $\mu$.  $T_S$ is also normalized with respect to its maximum value; from Fig. \ref{IFFT_Tslab} it can be noticed that temperature behaves as a damped wave propagating in time through the slab; moreover, as the value of  $\tau_{_T}$ is increased with respect to the laser pulse time $\tau_p$ the damping also increases. 
For instance, comparing the black and blue curve, at the first maximum value, temperature is around five times smaller for the larger thermal lag time. 

Before proceeding to analyze the acoustic pressure in frequency domain induced by DPL heat equation, it is important to notice that each of its corresponding differential equations given in Eqs. \eqref{1DPressure}, have a source term proportional to the second-order time derivative of temperature in the corresponding region, being the slab region $S$ where the heating process and acoustic waves generation occurs. 

\begin{figure}[ht]
  \centering
  \subfigure[]{
    \includegraphics[width=0.4\textwidth]{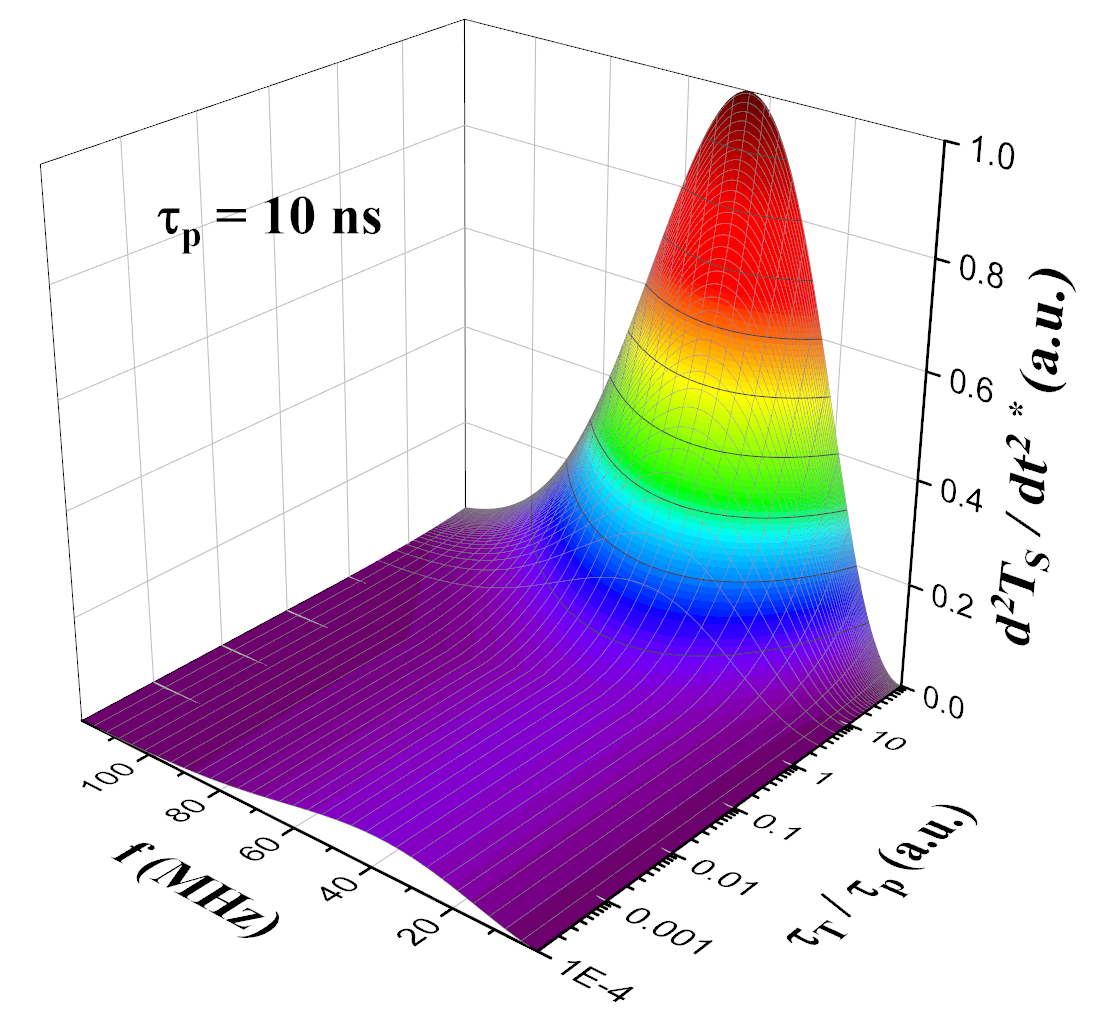}
    \label{fig:3DD2Temp}
  }
  \subfigure[]{
    \includegraphics[width=0.4\textwidth]{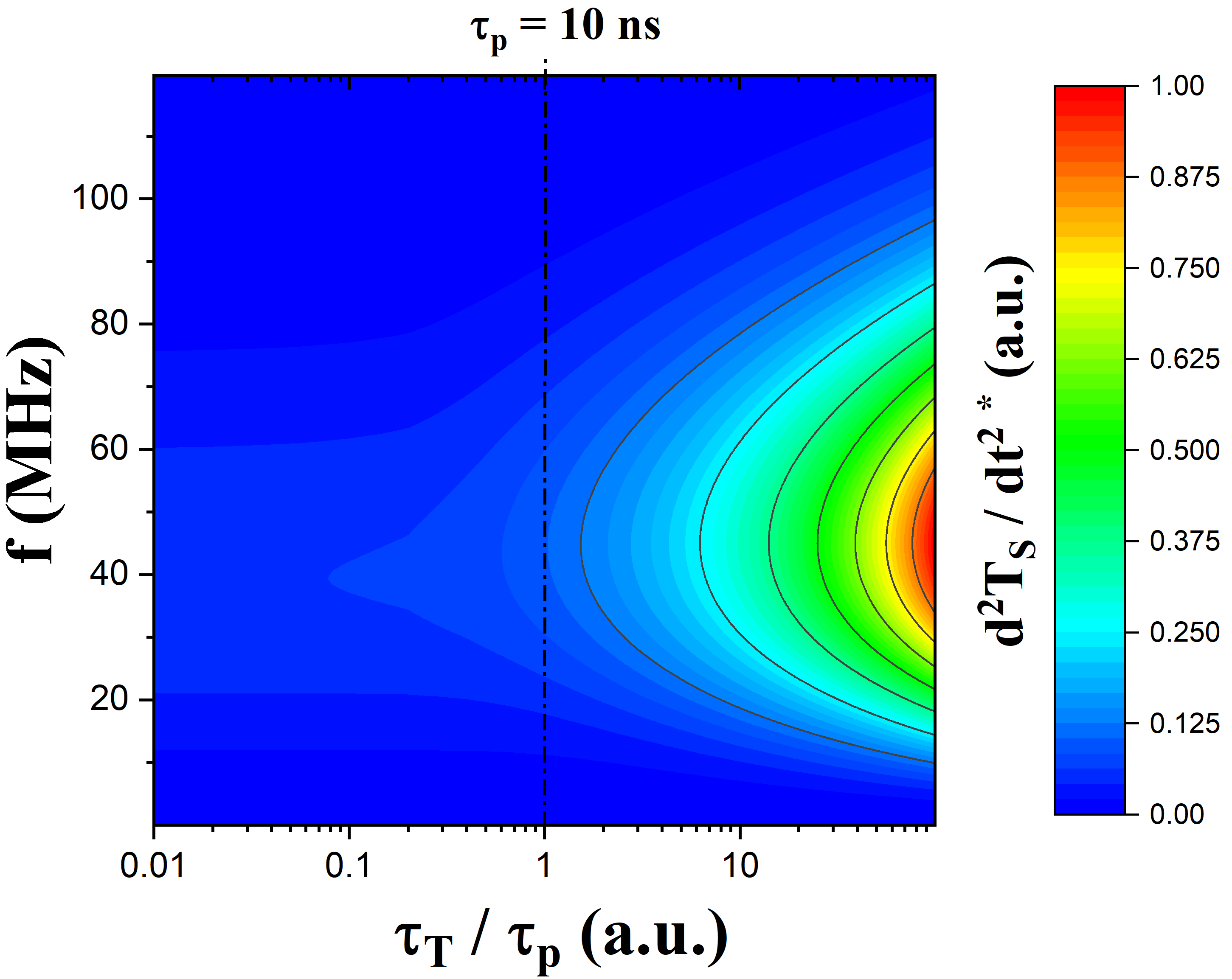}
    \label{fig:D2TempCon}
  }
  \\
  \hrulefill \\
  \caption{Normalized plots for the amplitude of the second-order time derivative of DPL temperature $\hat{T}_S(0,f;\tau_{_T})$. In \ref{fig:3DD2Temp} the 3D plot is presented, with $\tau_{_T}$ axis normalized with respect to a laser pulse time of $\tau_p = 10$ ns; for $\tau_{_T} < \tau_p$ second-order time derivative is well behaved; however once $\tau_{_T} \ge \tau_p$ a sharp increase appears in $d_{tt} \hat{T_S}$. In \ref{fig:D2TempCon} the corresponding contour plot is presented, divided in two regions by the dashed-dot line $\tau_{_T} = \tau_p$. }
  \label{fig:fig3}
\end{figure}

In the frequency domain, second-order time derivative operator is the reduced to $-\omega^2$; the corresponding source term in $S$ of acoustic pressure in this space is then proportional to $-\omega^2 \hat{T}_S (z,\omega;\tau_{_T})$. In Fig. \ref{fig:fig3} normalized plots for the second-order time derivative of temperature as a function of $f$ and $\tau_{_T}$ are presented.

As in the case of temperature, Fig. \ref{fig:fig3}, plots $\tau_{_T}$ normalized with respect to $\tau_p$ in order to compare how second-order time derivative is modified as $\tau_{_T}\to\tau_p$. In Fig. \ref{fig:3DD2Temp} the 3D plot is presented; a clear separation in two regimes is evident; for $\tau_{_T} < \tau_p$, this is a well behaved smooth function, however, for the interval $\tau_{_T} \ge \tau_p$ a sharp increase appears around $20\sim 80$ MHz in frequency. This is more evident in the corresponding contour plot given in Fig. \ref{fig:D2TempCon}; with the line $\tau_{_T} = \tau_p$ separating both aforementioned regions. A remarkable feature of this plot is how a small increase appears at $\tau_{_T} \approx 0.1 \ \tau_p$ centered around 40 MHz, becoming a steep increase as $\tau_{_T}\ge\tau_p$.

From these plots, it can be concluded that given the observed behavior in frequency domain of the heat source of wave equations for pressure, i.e., second-order derivative with respect to time ( or in frequency domain, $-\omega^2\hat{T}_S$), behavior of both $\hat{P}_F(z,\omega;\tau_{_T})$ and $\hat{P}_B(z,\omega;\tau_{_T})$ could be strongly modified for $\tau_{_T} \ge 0.1 \ \tau_p$. We will discuss this effect in the following. In this regard, the corresponding pressures are calculated at 1 nm away from the corresponding face of the slab.  

\subsection{Acoustic pressure}

\begin{figure}[ht]
  \centering
  \subfigure[]{
    \includegraphics[width=0.4\textwidth]{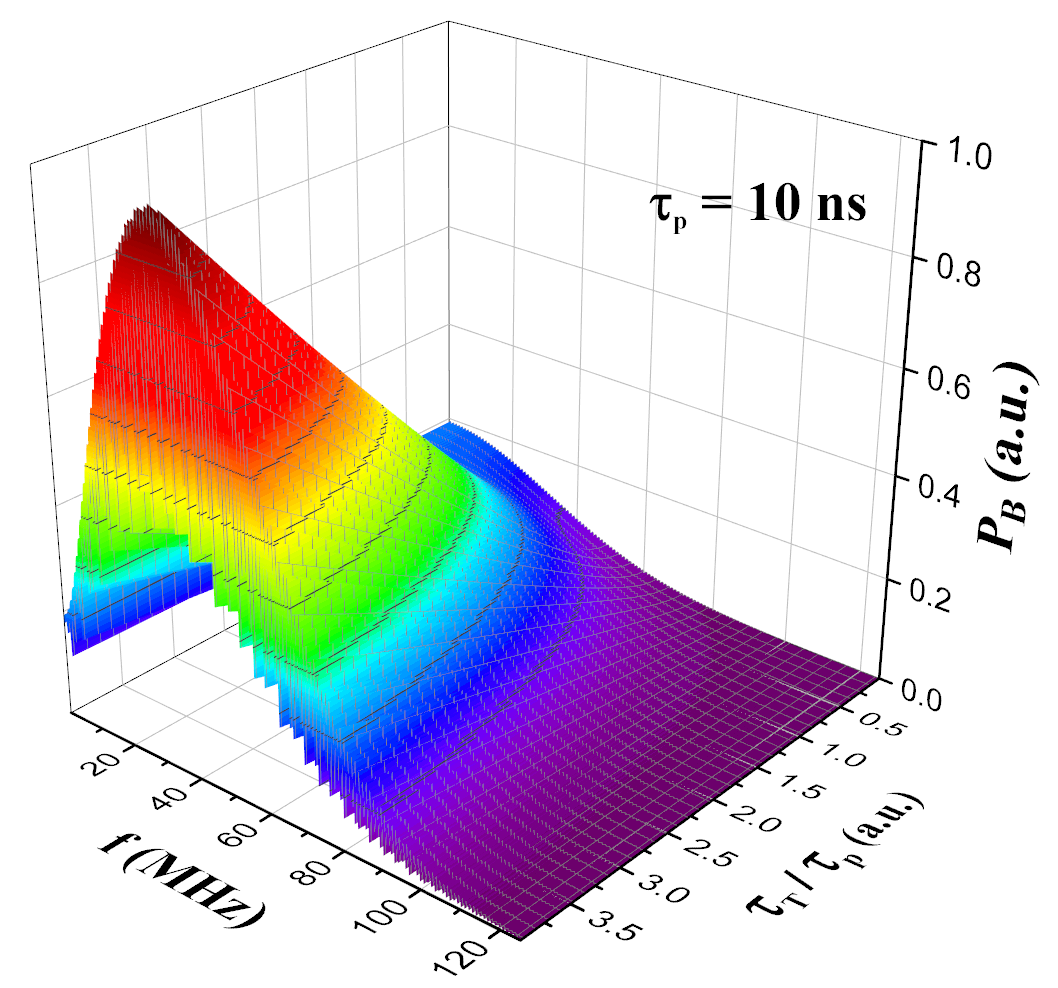}
    \label{fig:P_backward_3D}
  }
  \subfigure[]{
    \includegraphics[width=0.4\textwidth]{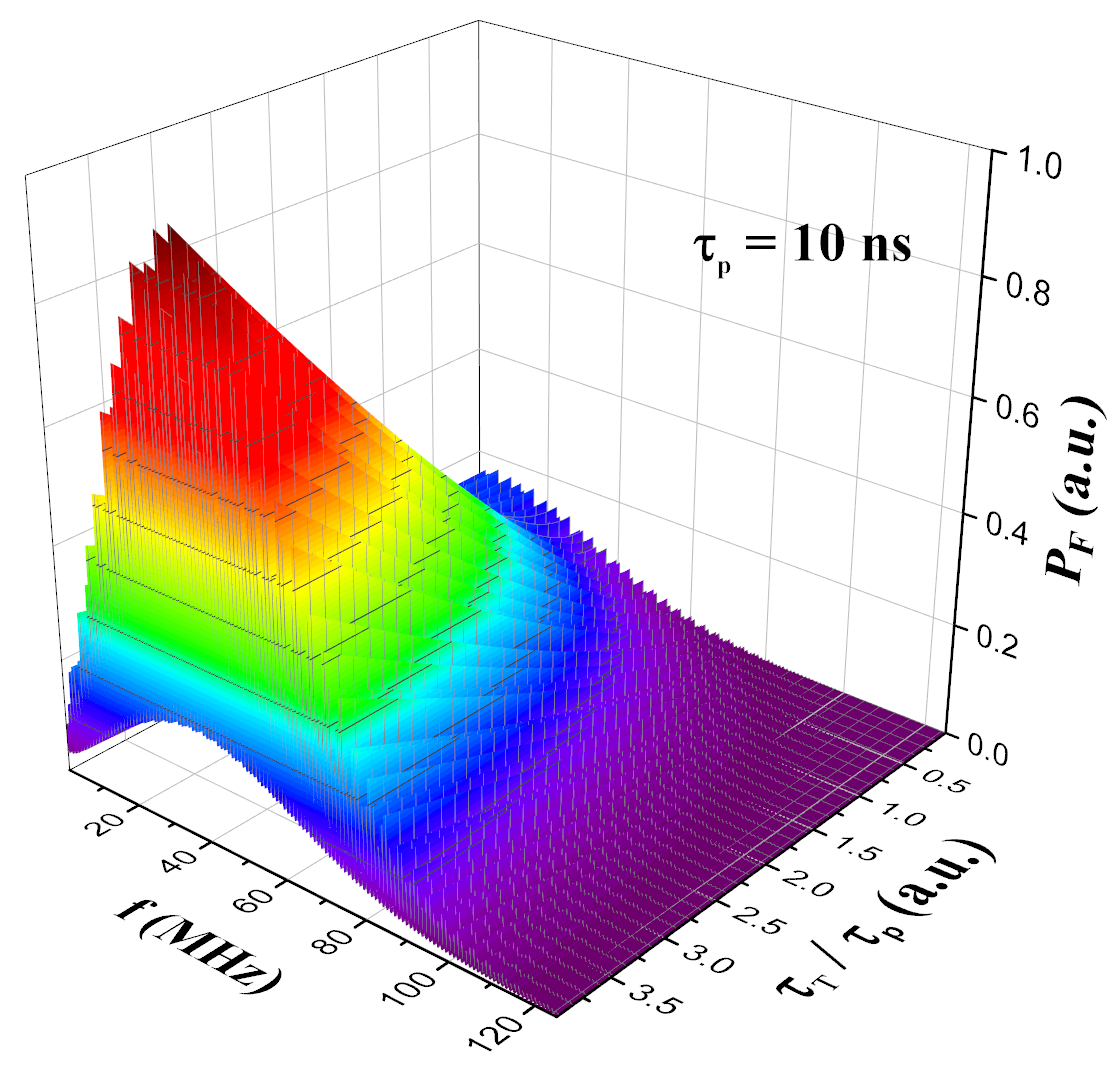}
    \label{fig:P_forward_3D}
  }
  \\
  \hrulefill \\
  \caption{3D plots for the amplitude of the pressure in frequency domain, normalized with respect to their corresponding maximum value. The pressure $\hat{P}_B$ for the region B is presented in (a) and the corresponding $\hat{P}_F$ in the region F is presented in (b). The characteristic structure related with the multiple acoustic reflections appears for both $\hat{P}_B$ and $\hat{P}_F$. It also can be noticed that as $\tau_{_T} \to \tau_p$, both pressures are modified, with an almost linearly increasing as $\tau_{_T} \ge \tau_p$ for a laser pulse time of 10 ns.}
  \label{fig:fig4}
\end{figure}

In Fig. \ref{fig:fig4} plots of normalized acoustic pressure  are presented as a function of frequency $f$ (in MHz) and thermal lag time $\tau_ {_T}$ also normalized with respect of the laser pulse time $\tau_p$, for both B and F regions; in Fig. \ref{fig:P_backward_3D} the induced pressure in the region B is presented and in Fig. \ref{fig:P_forward_3D} the corresponding plot for acoustic pressure in the region F is also presented. Both plots exhibit the well known multiple reflections structure of longitudinal waves which are characteristic of PPA effect in frequency domain for solid samples immersed in a fluid. 

Another important feature of acoustic pressure in both F and B regions, is that for $\tau_{_T}\le\tau_p$, both plots have an amplitude which has a close resemblance to the resulting pressure calculated from Eqs. \eqref{eq1}, i.e. the standard PPA model; however, when $\tau_{_T} > \tau_p$ a steady increase in amplitude appears for both $\hat{P}_B$ and $\hat{P}_F$. Accordingly, for a larger thermal lag time, compared to the laser pulse time, acoustic longitudinal waves will be significantly greater when compared to PHE results.  

\begin{figure}[ht]
  \centering
  \subfigure[]{
    \includegraphics[width=0.4\textwidth]{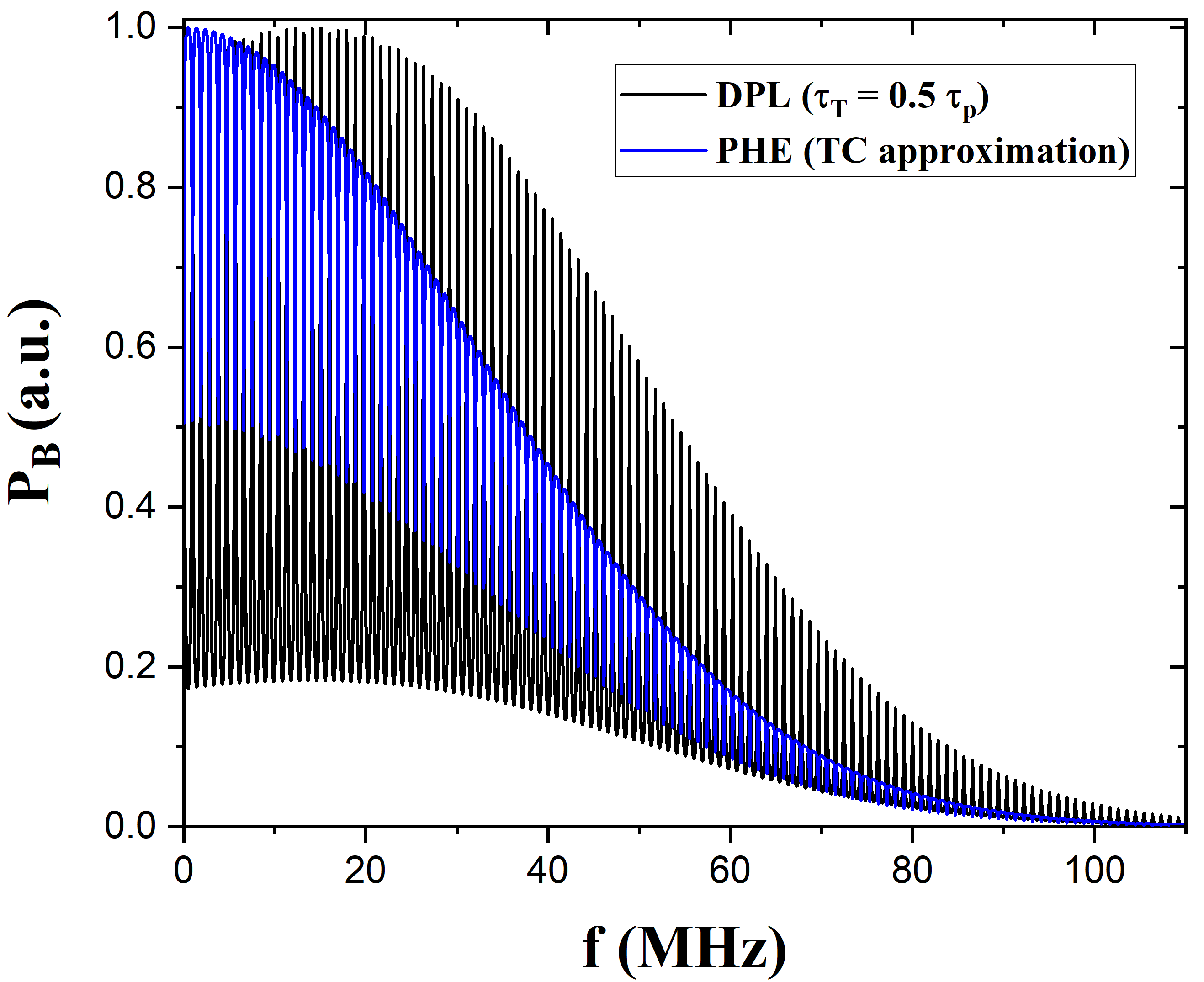}
    \label{fig:PB_DPL_vs_PHE}
  }
  \subfigure[]{
    \includegraphics[width=0.4\textwidth]{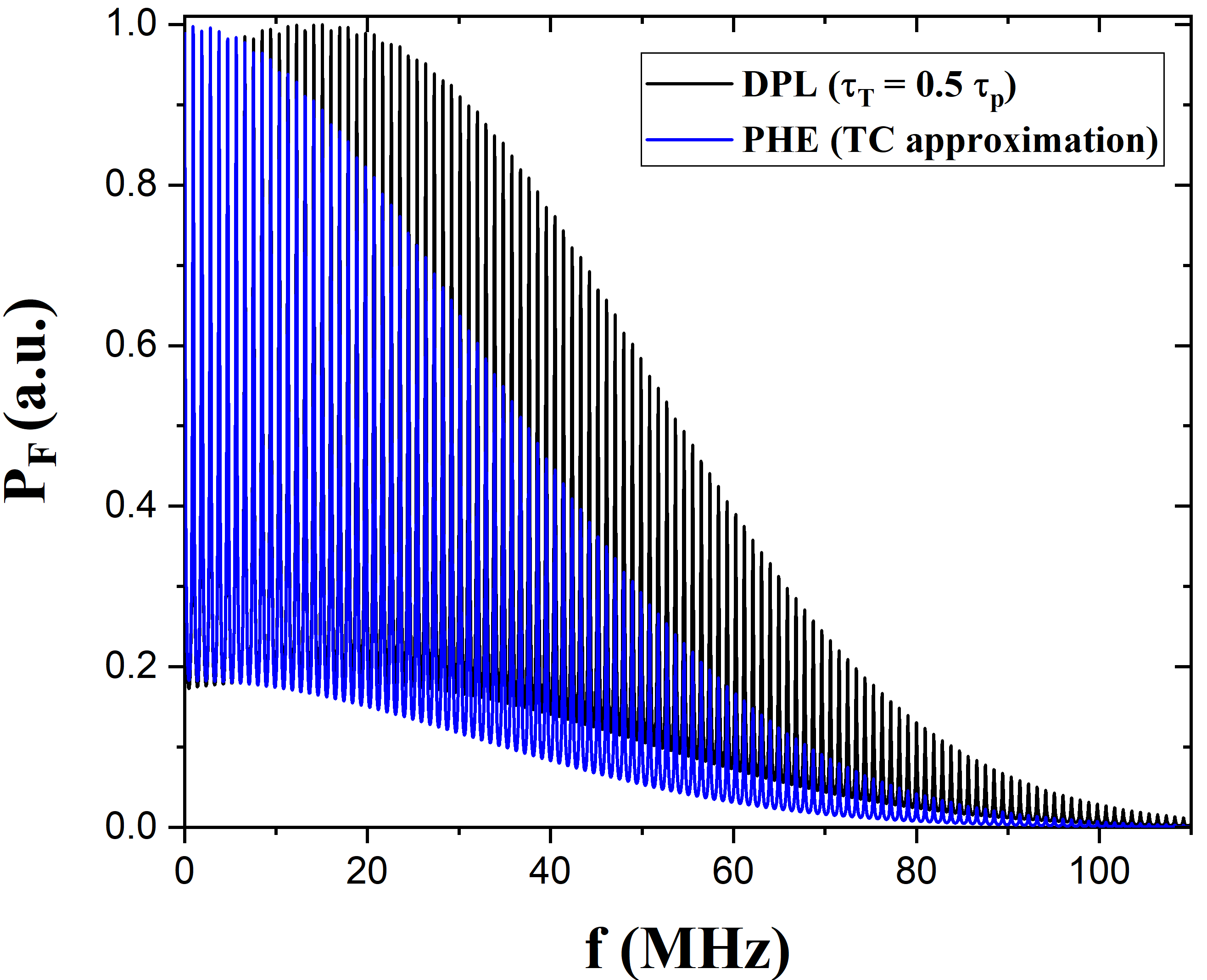}
    \label{fig:PF_DPL_vs_PHE}
  }
  \\
  \hrulefill \\
  \caption{Comparison of the normalized amplitude of the 2D acoustic pressure as a function of frequency, for the 1D three-layer problem calculated for the PPA model given the PHE with TC approximation (blue curve) and DPL heat equation (black curve), for a $3.4$ mm thick aluminum slab at $7.6$ mm away of each of the faces of the slab and  for $\tau_{_T}  = 0.5 \tau_p = 5 ns$. Where (a) is the pressure in B region and (b) in the F one. Both plots show a similar behavior, with a lag appearing in frequency between both models, with the PHE pressure always reaching its maximum at lower frequencies than the DPL one.}
  \label{fig:fig6}
\end{figure}

In Fig. \ref{fig:fig6} we present a direct comparison for the 1D three-layer problem of the normalized induced acoustic pressures in frequency domain for the PHE under TC approximation (blue curve) and for the DPL heat equation (black curve), for B and F regions in Fig. \ref{fig:PB_DPL_vs_PHE} and Fig. \ref{fig:PF_DPL_vs_PHE}, respectively. We have set the thermal lag time as $\tau_{_T} = 0.5 \tau_p = 5 ns$. Both models exhibit a very close resemblance regarding their shape and the multiple reflection structure.

From these figures, it can be noticed that for both B and F regions a lag in frequency between these models appears, with DPL pressure being shifted towards higher frequencies with respect to the PHE one, which always reach its maximum value near 0 MHz, and the DPL appearing around 20 MHz. Therefore, $\tau_{_T}$ could be considered as an adjusting parameter, in order to fit experimental results for PPA with this particular model. In the following the acoustic pressures generated by the DPL heat equation will be refereed as DPL--PPA 1D model. 



\section{\label{sec:VII} Experimental Procedure and DPL--PPA 1D model comparison}

\subsection{Experimental setup}

In this section a comparison between experimental data and the theoretical results for the acoustic longitudinal waves induced by the DPL heat equation, both in frequency and time domain, is presented and discussed, we also include analysis of the corresponding transference function between the modified PPA model and experimental data. For such analysis, the values used for thermomechanical properties of water and aluminium are presented  in Table \ref{tab:table2} of the supplemental material.


The considered experimental setup is presented in Ref. \cite{Rojas2024}. For which, a Q-Switch Nd: YAG pulse laser with a wavelength of 532 nm, average pulse duration of 10 ns, spot diameter of $\sim5$mm, and pulse energy of 26 mJ, is divided with a beam splitter 95/5. One beam passes through the window of a tank filled with water until it is focused ($\sim 1$ mm diameter spot), using a bi-convex lens on an aluminum plate, generating the LIU. The forward and backward LIU were detected using one piezodetector Panametrics M316-A, placed at  7.6 mm away from the aluminum slab, these signals were amplified using an amplifier. The amplifier was configured to work at 40MHz bandwidth. Experimental data were collected with an oscilloscope (9), which was triggered with the second pulsed beam on an IR nanosecond photo-detector and sent to a PC for processing. The dimensions (width×length×height) of the aluminum slab were 3.4 mm × 107.0 mm × 47.0 mm. 

\subsection{Frequency domain comparison}

\begin{figure}[ht]
  \centering
  \subfigure[]{
    \includegraphics[width=0.4\textwidth]{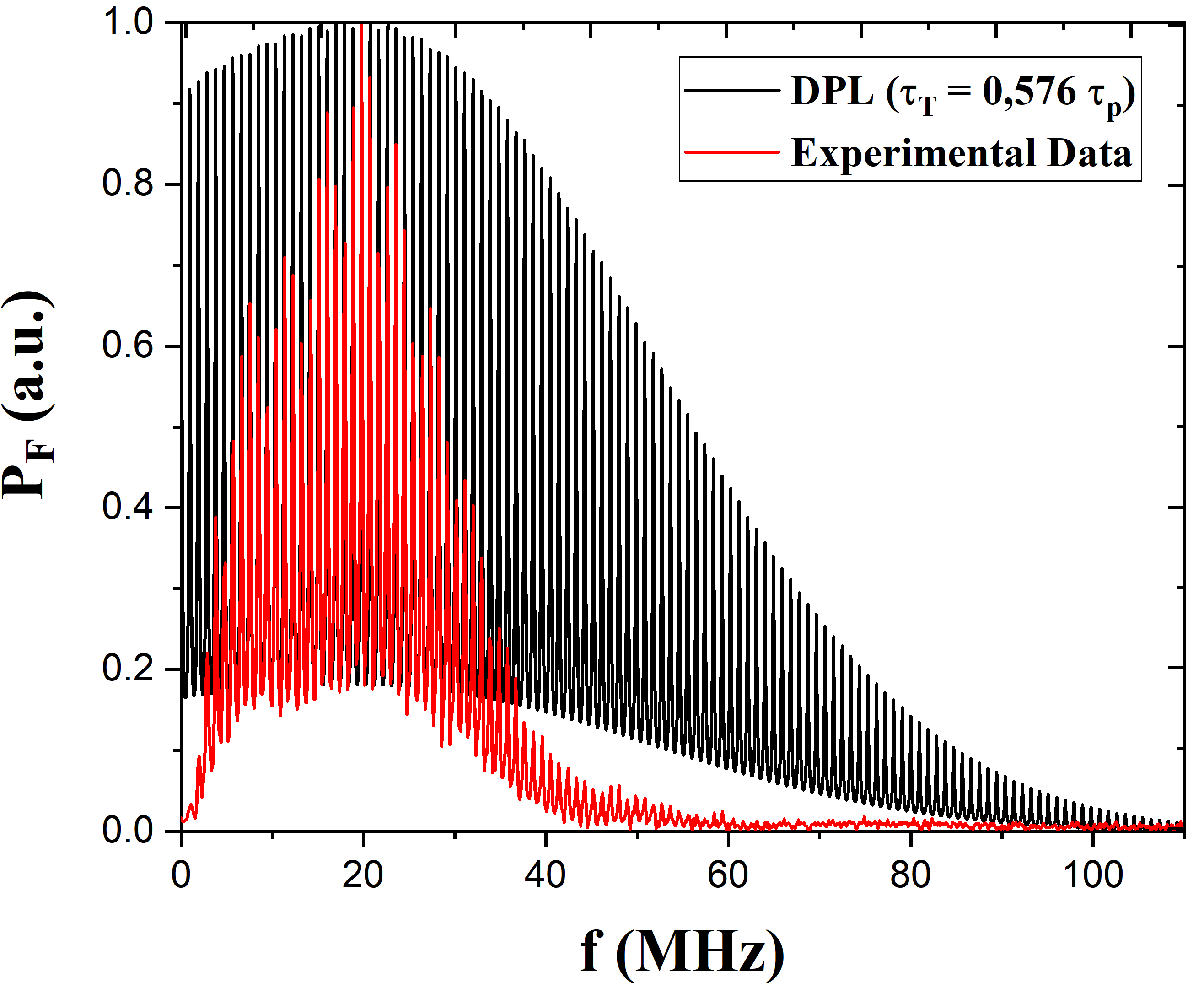}
    \label{fig:P_Forward_2D_full}
  }
  \subfigure[]{
    \includegraphics[width=0.4\textwidth]{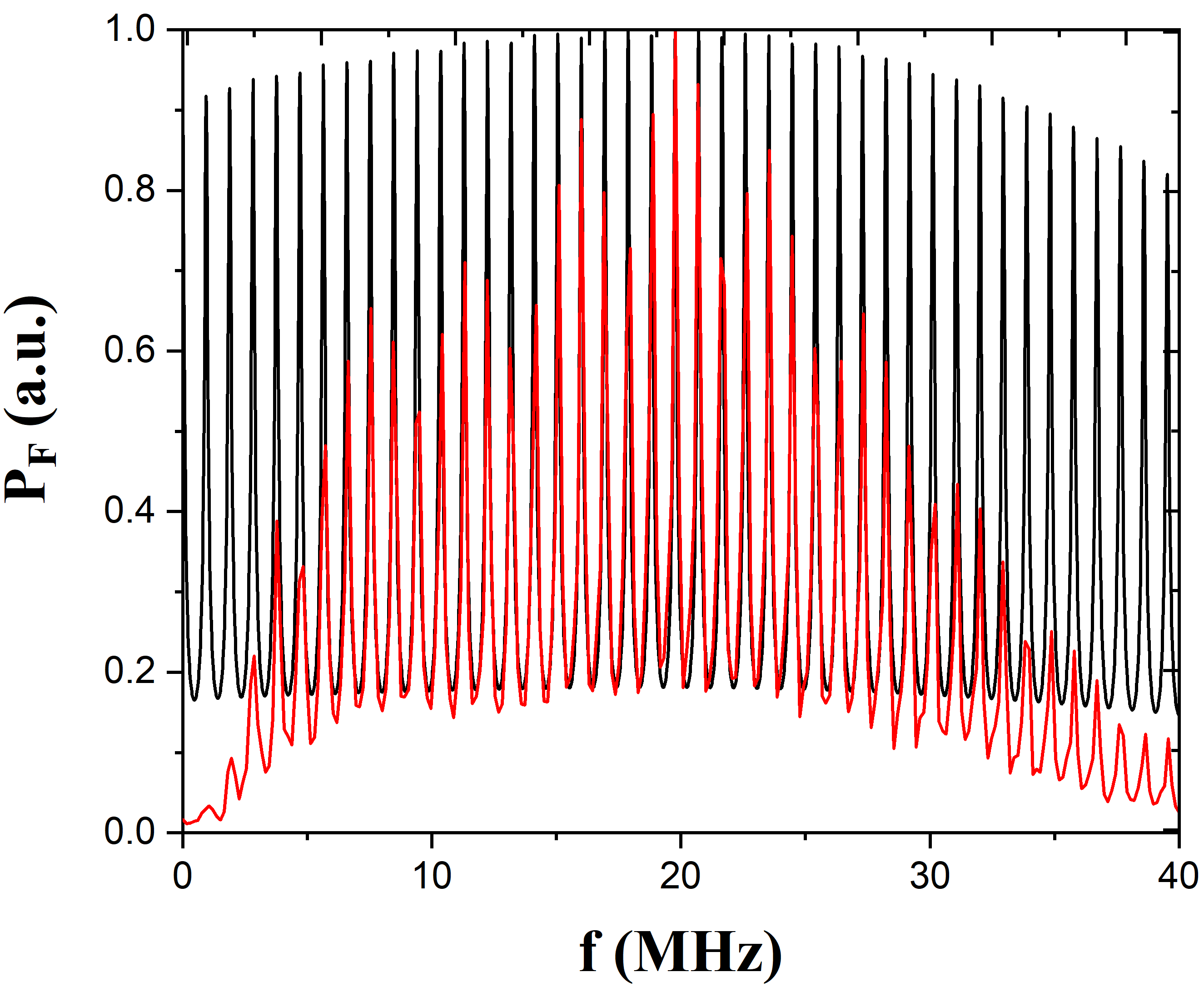}
    \label{fig:P_Forward_2D_zoom}
  }
  \\
  \hrulefill \\
  \caption{Normalized comparison in frequency domain between theoretical calculations, for the 1D three-layer DPL--PPA model (black line) given and experimental data (red line). Theoretical results are calculated at $7.6\times10^{-3}$ m away of the forward face of the aluminium slab. Full comparison is shown in Fig. \ref{fig:P_Forward_2D_full} and in Fig. \ref{fig:P_Forward_2D_zoom} this comparison is presented in the bandwidth of the amplifier (40 MHz).}
  \label{fig:fig8}
\end{figure}

In Fig. \ref{fig:fig8} a comparison in frequency domain, between an experimental PPA signal obtained in the F region with the above experimental setup and numerical results of the DPL--PPA 1D model given by Eqs. \eqref{Pressure_solutions} is presented; an aluminum slab of the same width and the same laser pulse time are considered in the theoretical model. Moreover, for this comparison, thermal lag time was set to $\tau_{T} =5.76$ ns. This value was numerically adjusted by considering the maximum value of frequency experimental data, which for this particular case is $f_{\text{max}} \approx 20$ MHz.

As noticed in Fig. \ref{fig:P_Forward_2D_full}, both signals have the same approximated shape; moreover, the multiple reflection structure of acoustic pressure in frequency domain is well reproduced by this model as shown in Fig. \ref{fig:P_Forward_2D_zoom}, which fits with experimental signal in the corresponding frequency values from 0 to 60 MHz. However, it is also noticeable that the experimental plot has a smaller bandwidth when compared to the signal calculated with the DPL--PPA 1D model. This difference in both signals could be a result of a loss of information produced by the piezo-detector response to the photoacoustic signal, which has not been considered in our model.

\begin{figure}[ht]
\centering
\includegraphics[width=0.4\textwidth]{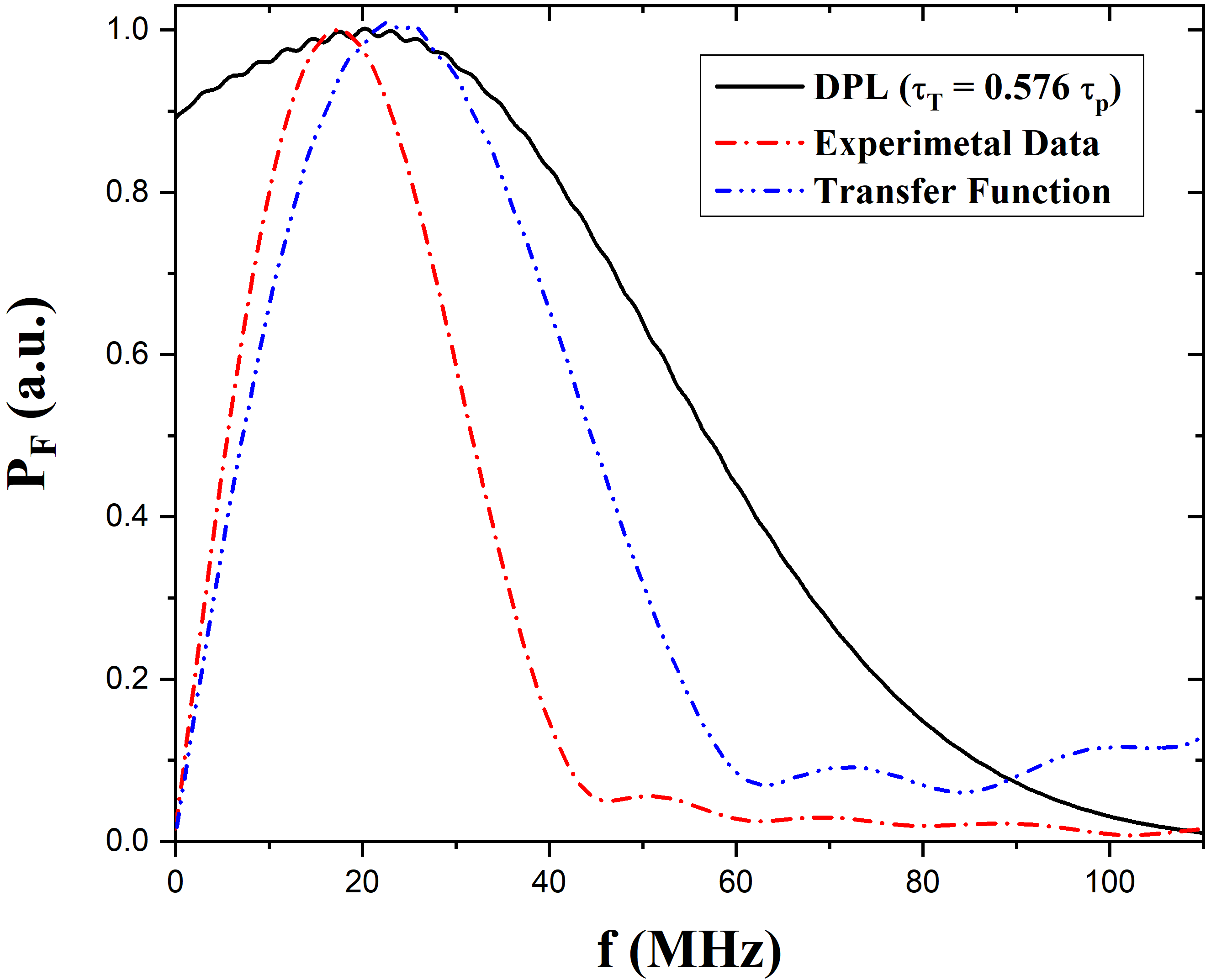}\\
\hrulefill \\
\caption{Transference function generated in frequency domain for experimental signal and the DPL--PPA 1D model for a thermal lag time of $\tau_{_T} = 5.76 $ ns. This function is constructed as the ratio between the envelope curves of experimental data and DPL--PPA 1D model. This function provide us with information regarding on the system response to different frequencies.}
\label{fig:fig9}
\end{figure}

Another important comparison in frequency domain between experimental signal and the DPL--PPA 1D model is presented in Fig. \ref{fig:fig9}, for which the corresponding transference function is calculated as the ratio between the envelope curves of experimental and the theoretical model. This function allow us to characterize the frequency response of the experimental system. Although the transfer function is directly influenced by experimental aspects such as the laser temporal profile and sensor bandwidth, it is precisely the analytical model which allows this information to be extracted. This is a simple example of how the analytical model can be used as a characterization tool for a photoacoustic experimental system. Therefore, from this function, we have found that the maximum response in frequency of this particular experimental setup is located around 22 MHz.

\subsection{Time domain comparison}

\begin{figure}[h]
  \centering
  \subfigure[]{
    \includegraphics[width=0.4\textwidth]{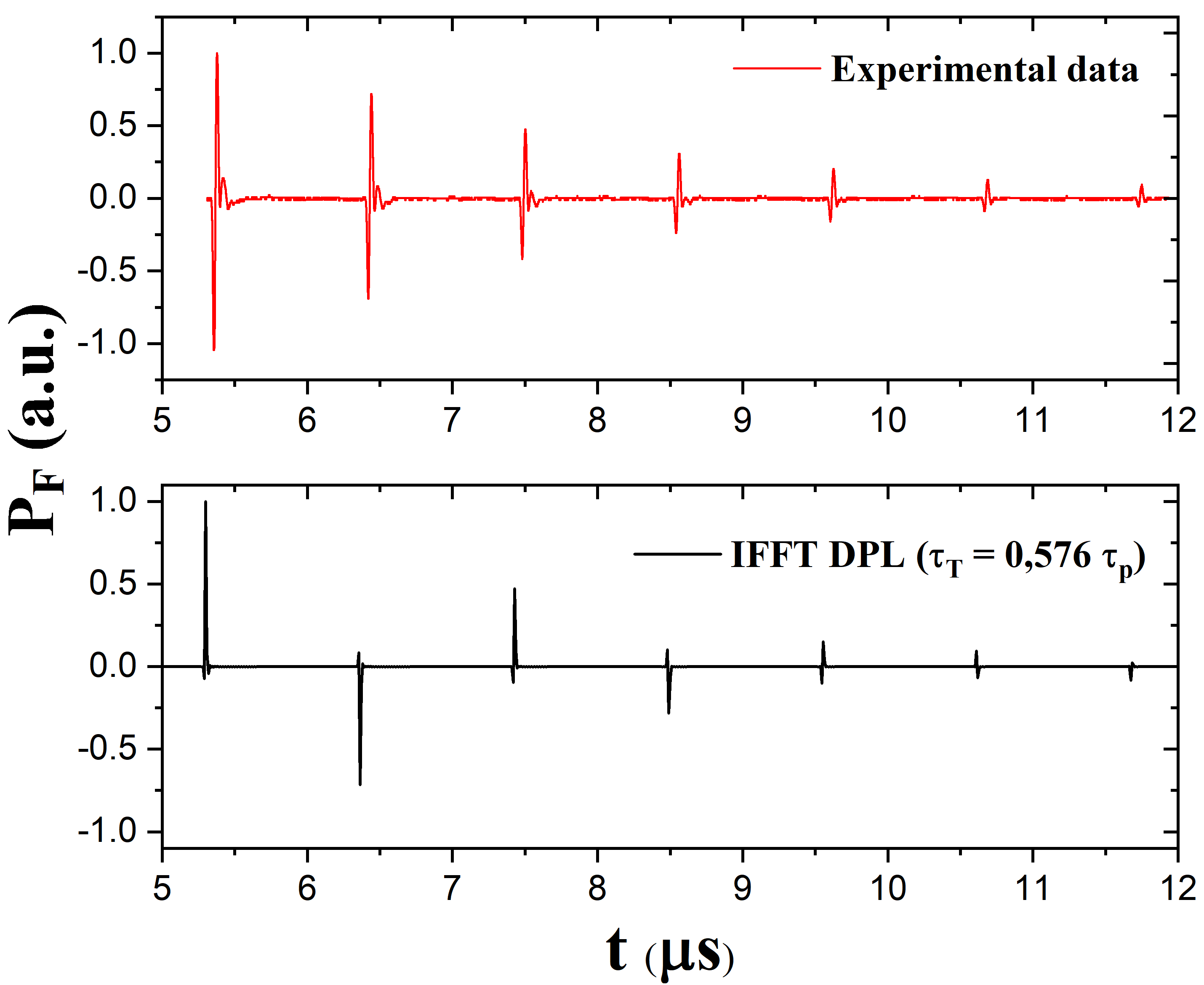}
    \label{fig:P_Forward_2D_time}
  }
  \subfigure[]{
    \includegraphics[width=0.4\textwidth]{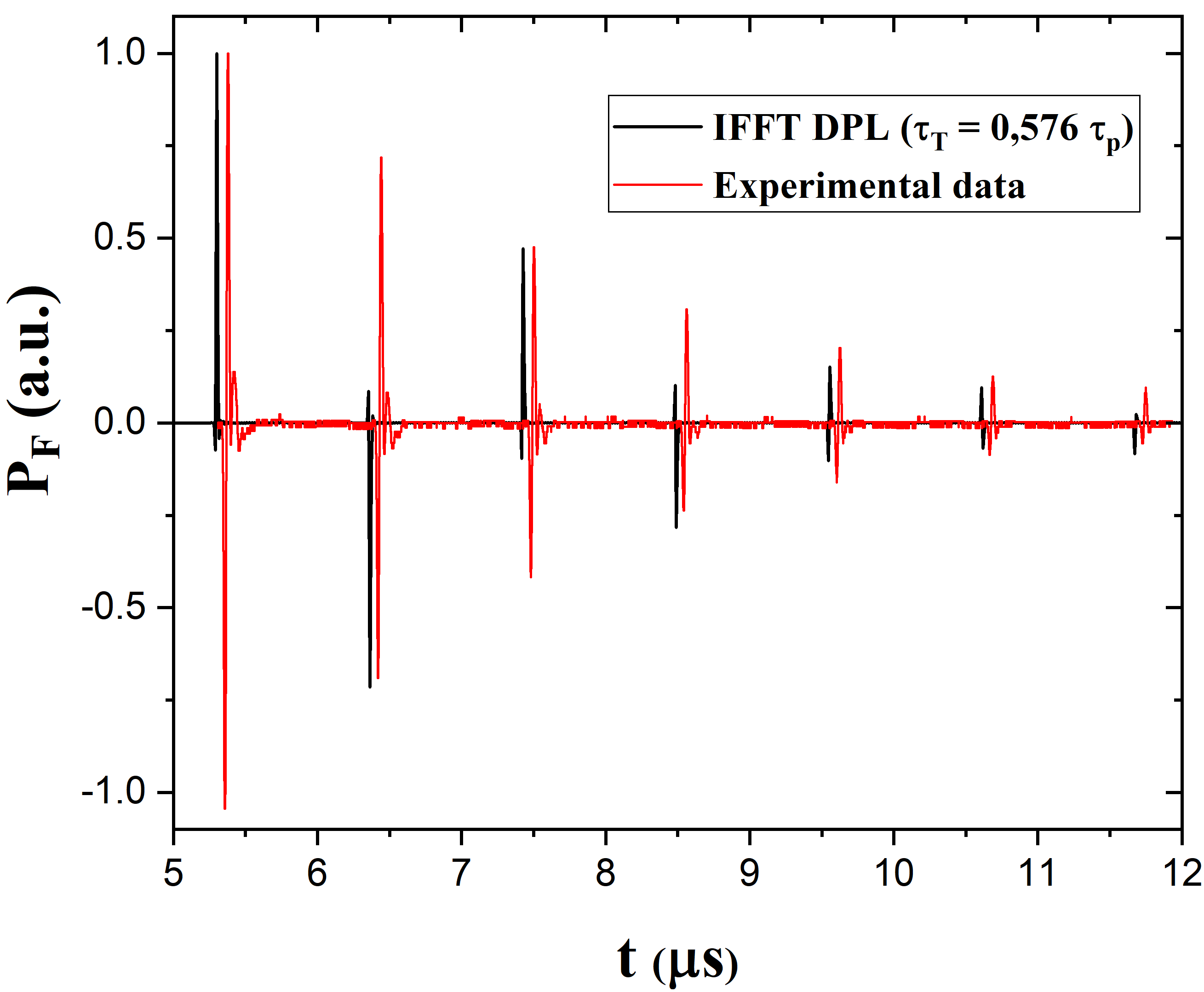}
    \label{fig:P_Forward_2D_comparison}
  }
  \\
  \hrulefill \\
  \caption{Comparison of the normalized acoustic pressure in forward region between experimental data and the DPL--PPA 1D model data in time domain for $\tau_{_T} = 5.76$. In (a) experimental signal and the real part of the numerical IFT DPL--PPA calculated signal for this system are presented separately; both curves exhibit the same general behavior, and each peak are equally separated. A direct comparison between both signals is presented in (b), with a slight shift around 80 ns, between the calculated signal and experimental signal is evident.}
  \label{fig:fig10}
\end{figure}

We are also interested to compare the DPL--PPA 1D model with experimental signal in time domain, which can be achieved, via an application of a numerical inverse Fourier transform on the DPL--PPA 1D acoustic pressure for the F region given by Eq. (\ref{Pressure_solutions}c). A comparison of the normalized acoustic pressures $P_F$ is then presented in Fig. \ref{fig:fig10}. As noticed, the DPL--PPA 1D model is capable of reproduce a structure with a close resemblance to the experimental signal. Additionally, the distance between each one of the peaks, which corresponds to the multiple reflections of the acoustic waves between the slab and the sensor, is approximately equal with both normalized amplitudes also present a similar behavior.

Direct comparison of both signals in time domain is presented in Fig. \ref{fig:P_Forward_2D_comparison}, for which the data of DPL--PPA 1D model appears slightly earlier in time with respect to experimental data, the interval of time between each of the corresponding peaks on the signals is $\Delta t = t_{_\text{EXP}} - t_{_\text{DPL}} \approx 80$ ns. At this moment it is not clear why this delay appears, however, it could be related with the electronic processing time of experimental data, or with a sensor response time delay, given that our model do not consider such effects. In a future work we will study this $\Delta t$ for different sensors and samples of different materials in order to explore this effect and how can it be considered in the DPL--PPA 1D model. 

\subsection{Comparison for different values of $\tau_{_T}$}

\begin{figure}[ht]
  \centering
  \subfigure[]{
    \includegraphics[width=0.4\textwidth]{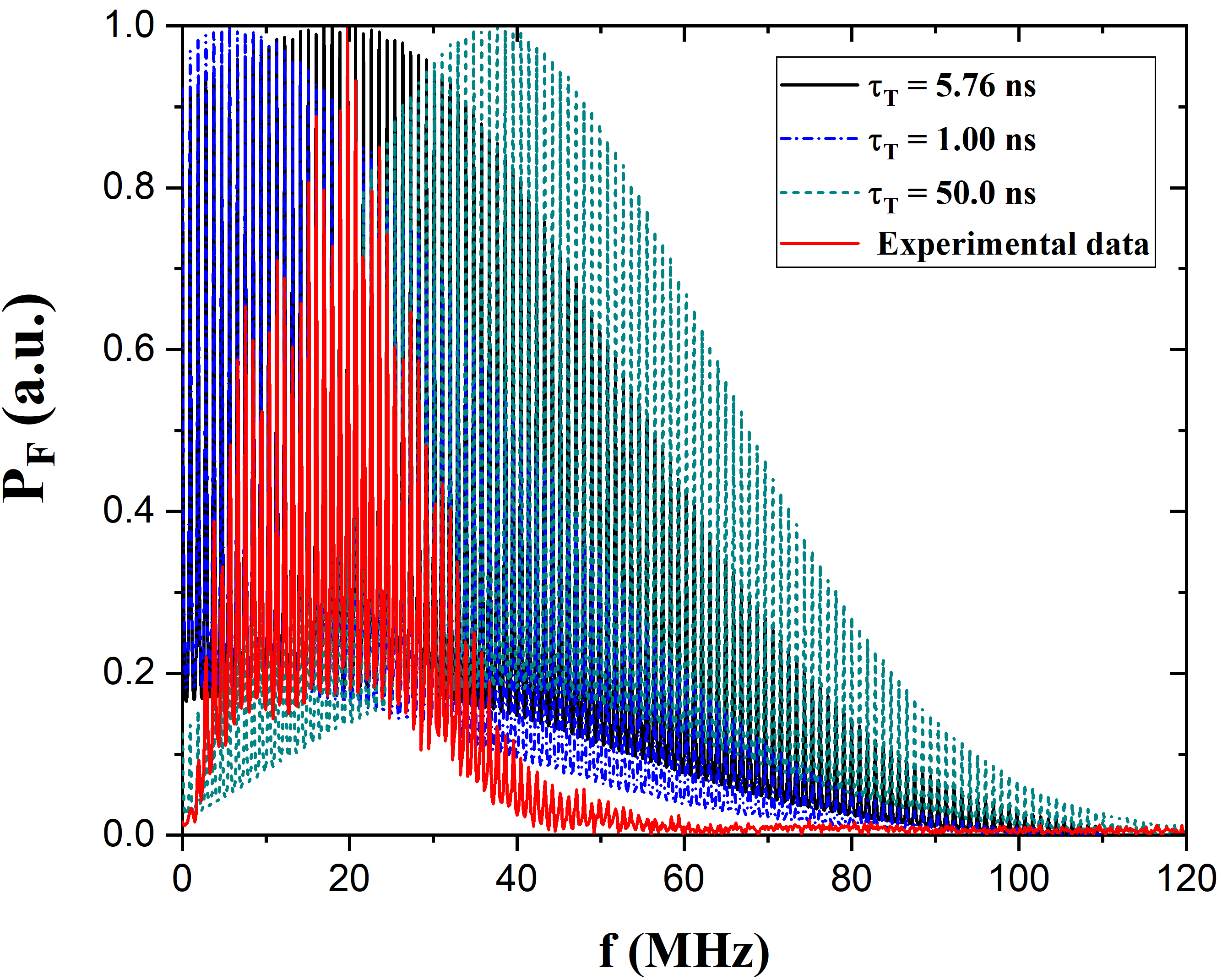}
    \label{fig:P_Forward_2D_tau_f}
  }
  \subfigure[]{
    \includegraphics[width=0.4\textwidth]{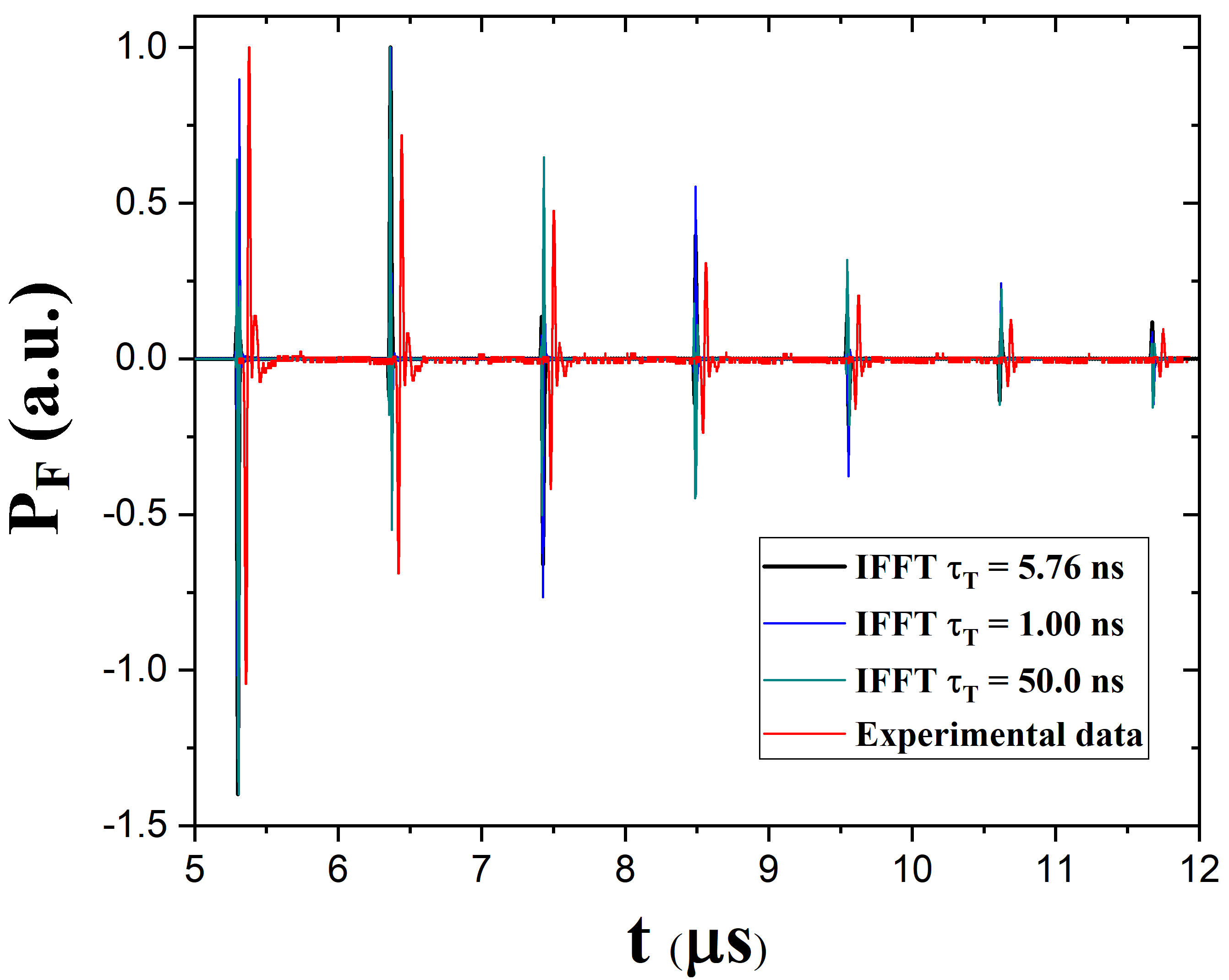}
    \label{fig:P_Forward_2D_tau_t}
  }
  \\
  \hrulefill \\
  \caption{Comparison in frequency and time domains of the normalized acoustic pressure in the forward region for different values for $\tau_{_T}$ and experimental signal (red curve), with the blue curve standing for $\tau_{_T}= 1$ ns, the black curve, $\tau_{_T}= 5.76$ ns, the green one corresponds to $\tau_{_T}= 50$ ns and the red curve stands for experimental results.}
  \label{fig:fig11}
\end{figure}

Finally, we present a comparison for F region between the experimental signal and the DPL--PPA 1D model for different values of $\tau_{_T}$, namely, 1 ns, 5.76 ns and 50 ns, for both frequency and time domain,  presented in Fig. \ref{fig:fig11}. Frequency domain is presented in Fig. \ref{fig:P_Forward_2D_tau_f}, in this plot we show how for different values of $\tau_{_T}$ in the DPL--PPA 1D model, the frequency value at which the acoustic pressure reach its maximum value is modified. From this figure, it is expected that the PPA pulse have a narrower frequency content; however, in time domain presented in Fig. \ref{fig:P_Forward_2D_tau_t} due to the scale of the plot in time domain and the slight variation on their frequency content, this effect is barely noticeable.

\section{\label{sec:VII} Conclusions and perspectives}

In this work we have presented a new theoretical model for the 1D pulsed photoacoustic effect constructed by considering a generalization to the classical Fourier heat conduction model, introducing a delay both in heat flux and thermal diffusion generated in a system which  transforms optical energy into heat. This is achieved by the Dual-Phase-lag heat conduction model, which is characterized by two parameters, $\tau_q$ and $\tau_{_T}$. The first one is related with a time lag in heat flux when the sample is heated; theoretically, $\tau_q$ is related with  perturbations in the density of the phonon gas, i.e., a wavelike propagation of heat. $\tau_{_T}$ corresponds to a lag on the thermal response of the heated sample. To our knowledge it has not been related yet with any micro/mesoscopic phenomena. Therefore, in this work we have
made two assumptions regarding this parameter; for the sake of simplicity, in the system under study $\tau_{_T}$ is assumed to be the same regardless of the material. The second assumption is that $\tau_{_T}$ is a free parameter in our model, to be adjusted to accurately represent experimental signals.

We were able to solve this 1D heat equation in frequency domain, for a three-layer system with boundary conditions, for which the DPL heat equation reduces to a second order ordinary differential equation. This solution is then considered to be the acoustic source term that generates the pulsed photoacoustic effect on a sample slab which is modeled via the wave equation given by Eq. (\ref{eq1}b). This new DPL--PPA model allow us to find analytical expressions in frequency domain for the acoustic pressure for the 1D boundary three-layer problem. We were able to compare these theoretical results with experimental data, showing that if the thermal lag time is set at $\tau_{_T} = 5.76$ ns, both experimental and numerical data have a very close resemblance in their structure in the frequency domain with an narrower bandwidth in the experimental case; probably due to the non-ideal sensor response. In time domain we have found that, via a numerical inverse Fourier transform, theoretical acoustic pressure closely resembles experimental data, with the exception of a slight time delay, which could be the result of a delay in sensor response. 

It is clear that this research represents only the first step in the study of the DPL-PPA model, and more work both theoretical and experimental  is needed in order to accurately model the laser induced ultrasound for more general problems; besides additional research is needed to be able to construct analytical expressions for the acoustic pressure in time domain instead of frequency. Additionally, from the experimental perspective, experimental setups for different sensors and samples must be explored and compared with the DPL-PPA model to test its accuracy and try to explain some important aspects which are not yet clear; such as the $\Delta t$ appearing between theoretical calculation and experimental data. 

\section{Supplementary Material}
In the supplemental material the photothermoacoustic boundary conditions considered for the 1D three-layer problem are presented; we also present the explicit functional form of the undetermined coefficient for both the DPL thermal equation and the acoustic wave equation for the considered problem. Also, a table of the thermomechanical properties used in the numerical calculations presented in the manuscript is also given.

\section*{Acknowledgments}
L. F. Escamilla-Herrera, O. Medina-Cázares and J. E. Alba-Rosales acknowledge support from CONAHCyT (Consejo Nacional de Humanidades Ciencias y Tecnologías) through postdoctoral grants: Estancias Posdoctorales por México para la Formación y Consolidación de las y los Investigadores por México (CVU's: 230753, 241606, 419686  respectively). J. M. Derramadero-Dominguez acknowledge support from CONAHCyT  through the Master Scholarship (CVU: 1275184). G. Gutíerrez-Juárez acknowledge partial support from CONAHCyT grant CBF2023-2024-3038; also was partially supported by DAIP-Universidad de Guanajuato: CIIC Grant No. 209/2024.


\bibliographystyle{apsrev}
\bibliography{base_datos_PA}

\newpage

\end{document}